\documentclass[acmsmall]{acmart}

\usepackage{array}
\usepackage[normalem]{ulem}
\usepackage{xcolor}

\AtBeginDocument{%
  }

\setcopyright{acmlicensed}
\copyrightyear{2026}
\acmYear{2026}
\acmDOI{XXXXXXX.XXXXXXX}

\acmConference[CSCW '26]{ACM Conference on Computer-Supported Cooperative Work and Social Computing Conference}{Oct 10--14,2026}{Salt Lake City, Utah, USA}

\acmISBN{978-1-4503-XXXX-X/18/06}

\begin{document}

\title{\name{}: An Elementary Grammar for Sharing Expertise in Craft Workflows}

\author{Ritik Batra}
\affiliation{%
  \institution{Cornell Tech}
  \city{New York}
  \country{USA}}
\email{rb887@cornell.edu}

\author{Lydia Kim}
\affiliation{%
  \institution{Cornell Tech}
  \city{New York}
  \country{USA}}
\email{lk569@cornell.edu}

\author{Ilan Mandel}
\affiliation{%
  \institution{Cornell Tech}
  \city{New York}
  \country{USA}}
\email{im334@cornell.edu}

\author{Amritansh Kwatra}
\affiliation{%
  \institution{Cornell Tech}
  \city{New York}
  \country{USA}}
\email{ak2244@cornell.edu}

\author{Jane L. E}
\affiliation{
  \institution{Stanford University}
  \city{Stanford}
  \country{USA}}
\email{ejane@stanford.edu}

\author{Steven Jackson}
\affiliation{
  \institution{Cornell University}
  \city{Ithaca}
  \country{USA}}
\email{sjj54@cornell.edu}

\author{Thijs Roumen}
\affiliation{%
  \institution{Cornell Tech}
  \city{New York}
  \country{USA}}
\email{thijs.roumen@cornell.edu}

\renewcommand{\shortauthors}{Batra et al.}

\newcommand{\scratch}[1]{\textcolor{red}{#1}}

\newcommand{\name}{\textit{(De)composing Craft}}
\newcommand{\grammar}{\textit{elementary grammar}}

\definecolor{niceorange}{RGB}{210,125,45}
\newcommand{\rrsep}[2]{\textcolor{black}{{#2}}}

\definecolor{niceblue}{RGB}{0,71,171}
\newcommand{\rrnov}[2]{\textcolor{black}{{#2}}}

\newcommand{\detailblocks}{\textsc{Granularity Shifts}} 
\newcommand{\reflective}{\textsc{Reflective Loops}}
\newcommand{\note}{\textsc{Note-to-Self}}
\newcommand{\external}{\textsc{External Links}}
\newcommand{\segments}{\textsc{Segments}}
\newcommand{\branches}{\textsc{Branches}}
\newcommand{\revisionloops}{\textsc{Revision Loops}}

\newcommand{\interface}{CraftLink}

\newcommand{\rb}[1]{\textcolor[RGB]{0,0,255}{[RB: #1]}}
\newcommand{\tr}[1]{\textcolor[RGB]{128,0,128}{[TR: #1]}}
\newcommand{\sj}[1]{\textcolor[RGB]{0,128,0}{[SJ: #1]}}
\newcommand{\je}[1]{\textcolor[RGB]{128,128,0}{[JE: #1]}}
\newcommand{\ak}[1]{\textcolor[RGB]{0,128,128}{[AK: #1]}}


\begin{abstract}

Craft practices rely on evolving archives of skill and knowledge\rrnov{,}{} developed through generations of craftspeople experimenting with designs, materials, and techniques. Better documentation of these practices enables the sharing of knowledge and expertise between sites and generations. \rrnov{However, most documentation focuses solely on the linear steps leading to final artifacts, neglecting the tacit knowledge necessary to improvise, or adapt workflows to meet the unique demands of each craft project.}{However, most documentation focuses on the linear steps leading to final artifacts, neglecting the distinct tacit knowledge, improvisational actions, and situated adaptations needed to meet the unique demands of each craft project.} This omission limits knowledge sharing and reduces craft to a mechanical endeavor, rather than a sophisticated \rrnov{}{and contextual} way of seeing, thinking, and doing. Drawing on expert interviews and literature from HCI, CSCW and the social sciences, we develop an \emph{elementary grammar} to document improvisational actions of real-world craft practices. \rrsep{We demonstrate the utility of this grammar with an interface called \emph{CraftLink} that can be used to analyze expert videos and semi-automatically generate documentation to convey material and contextual variations of craft practices.}{We demonstrate the utility of this grammar with \rrnov{an}{a MLLM-powered} interface called \emph{CraftLink} that can be used to analyze expert videos and generate documentation to share material and contextual variations of practices with other knowledgeable but non-master craftspeople.} Our user study with expert crocheters (N=7) \rrnov{using this interface }{}evaluates our grammar's effectiveness in capturing and sharing expert knowledge with other craftspeople, offering new pathways for computational systems to support collaborative archives of knowledge and practice \rrsep{within communities}{across time, space, and skill levels}. \rrnov{}{We conclude by showing how our grammar address four key tensions of the craft learning environment: personal and shareable documentation, fragmented and discoverable expertise, linear and iterative practices, and data privacy and ownership.}


\end{abstract}



\received{13 May 2025}
\received[revised]{13 January 2026}

\maketitle

\section{Introduction}

\begin{quote}
    \emph{[Craftsmanship] means simply workmanship using any kind of technique or apparatus, in which the quality of the result is not predetermined, but depends on the judgment, dexterity and care which the maker exercises as he works. The essential idea is that the quality of the result is continually at risk during the process of making; and so I shall call this kind of workmanship ``The workmanship of risk.''} --- David Pye~\cite{pye1995nature}
\end{quote}

\noindent Craft, \rrnov{defined by David Pye as the \emph{workmanship of risk},}{as defined by David Pye in the passage above,} relies on evolving archives of skill and knowledge, continually shaped by material engagement\rrnov{}{s} and emerging tools~\cite{ingold2013making}.
As contributors to this growing \rrnov{}{and collaborative} archive, craftspeople experiment with and share their designs, materials, and techniques with others~\cite{torrey2009learning}.
This \rrnov{knowledge }{}exchange is essential to craft's dynamism: its perpetual practices of adjustment, improvisation, and reinvention \rrnov{keep it current}{that establish craft as a living and continually evolving field of practice}~\cite{cheatle2015digital, cheatle2023re, apaydin2020dowry}.
\rrnov{Expert}{Through these practices, expert} craftspeople develop not only procedural fluency, but also a \rrnov{nuanced, tacit}{supple and nuanced} understanding of their craft --- situated knowledge that is difficult to articulate yet central to their practice~\cite{polanyi1967tacit}.
\rrnov{}{In many fields and settings, this learning is achieved through extended and embodied co-presence, often through formal and informal apprentice relations organized into wider communities of practice~\cite{cheatle2023re, lave1991situated}.}
\rrnov{}{
After long periods of practice~\cite{ericsson1993role}, expert craftspeople develop a deep understanding of the structure or \emph{rules} of their craft --- the techniques, processes, and procedures by which high-quality objects are produced.
But they also understand the conditions under which rules are to be amended or broken and the interpretive freedom needed to navigate unique conditions, materials and contexts.
}

\rrnov{However, a key challenge}{A key challenge however} arises when craft knowledge is shared across distances (remotely) and time (asynchronously).
\rrnov{Craftspeople}{In the absence of face-to-face interactions, craftspeople often} rely on static documentation, such as written instructions, video tutorials, and design files, to share knowledge.
\rrsep{While these formats convey formal specifications and mechanical procedures, they fail to transmit the situated and tacit knowledge needed to navigate dynamic or uncertain contexts~\cite{suchman1987plans}.}{While these formats convey mechanical procedures that may be useful for novices learning fundamentals, they fail to transmit the tacit knowledge, improvisational actions, and situated adaptations needed for other knowledgeable but non-master craftspeople to navigate similar contexts~\cite{suchman1987plans}.}
\rrnov{After long periods of practice~\cite{ericsson1993role}, expert craftspeople develop a deep understanding of the structure or \emph{rules} of their craft --- the techniques, processes, and procedures by which high-quality objects are produced.
But they also understand the conditions under which rules are to be amended or broken and the interpretive freedom needed to navigate unique conditions, materials and contexts.}{}
These deviations are not random; \citet{goodwin2015professional} shows how situated actions stem from \emph{professional vision}, where experts draw on experiential knowledge to interpret and respond to dynamic contexts.
With this ability to discern patterns and contextual cues, experts adapt their responses fluidly in shared but distinctive ways through \emph{improvisation}~\cite{berliner2009thinking}.
Capturing improvisations is challenging, as it requires representing implicit decisions and actions as structured and navigable documentation artifacts~\cite{fischer2005knowledge, suthers2003deictic} that support collaboration and the sharing of situated and tacit knowledge~\cite{ackerman2013sharing}.

\begin{figure}[h]
  \centering
  \includegraphics[width=\linewidth]{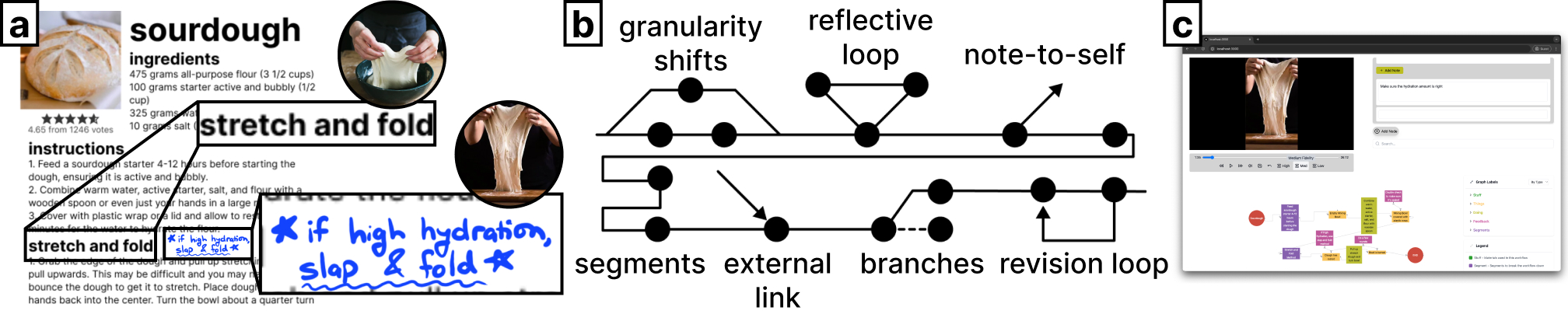}
  \caption{
  (a) Improvisations in craft are documented as annotations, shown in this sourdough recipe.
  However, sharing \rrnov{situated and tacit knowledge}{tacit knowledge, improvisational actions, and situated adaptations} is crucial to advance craft knowledge, \rrnov{here shown}{shown here} by different kneading techniques of \emph{stretch and fold} compared to \emph{slap and fold}.
  (b) We developed an \grammar{} composed of seven patterns occurring in craft workflows, to share \rrnov{knowledge}{expertise} \rrnov{between craftspeople}{within craft communities}.
  (c) We demonstrate the grammar in practice by designing an interface called \emph{\interface{}} that enables craftspeople to translate unstructured craft videos to shareable documentation artifacts, based on our grammar.
  }
  \label{fig:teaser}
  \Description{}
\end{figure}

Figure~\ref{fig:teaser}\rrnov{}{a} illustrates how tacit knowledge influences the interpretation of craft documentation.
\rrnov{The}{An} expert baker \rrnov{views}{may view} the recipe as a starting point, not a strictly determinate sequence (Figure~\ref{fig:teaser}a).
Substantial experience allows them to recognize that each step represents an action within a larger workflow, and that adjustments may be necessary, as shown by their annotation to use the \emph{slap and fold} kneading technique when the dough has a higher hydration as opposed to the original \emph{stretch and fold}.
\rrnov{Conversely, another baker with less familiarity with sourdough may approach the recipe as immutable truth, following each instruction without grasping the underlying rationale. Much like in jazz, where musicians improvise based on a shared understanding of standards, scales and rhythms~\cite{berliner2009thinking}, craft knowledge thrives in collaborative environments where understanding is not fixed but evolves through communal engagement and sharing improvisational practices~\cite{kang2021techarttheory}.}{Conversely, another baker exploring alternative methods of making sourdough may approach the recipe as a fixed structure, following each instruction without grasping the underlying rationale. Much like in jazz, where musicians improvise based on a shared understanding of standards, scales and rhythms~\cite{berliner2009thinking}, craft knowledge thrives in collaborative communities where understanding evolves through experimentation, variation, and the sharing of processes~\cite{kang2021techarttheory}.}

Andrew Abbott has demonstrated how these actions are \emph{rule-based}, not \emph{rule-bound}~\cite{abbott2014system}.\rrsep{This distinction marks the difference between a practice of craft that is rote, mechanical and static and one that is live, dynamic, and evolving. This distinction renders existing static methods of documentation, like the recipe shown in Figure~\ref{fig:teaser}a, ineffective.}{}
\rrnov{\emph{Rule-bound} approaches communicate the explicit rules of craft practice but cannot capture the nuanced improvisation and adjustment that experts perform in everyday practice.
They can show practice in the \emph{standard case}, but rarely the myriad deviations and adjustments to workflows that mark the practice and imagination of the expert craftsperson (and ultimately produce objects or things of distinctive quality or character).}{In \emph{rule-bound} approaches, practice is fully captured and `exhausted' by rules, with no space for variation or deviation from the set path that rules prescribe.}
\rrnov{}{A \emph{rule-based} approach, by contrast, centers the craftsperson's ability to work within established rules while knowing when to break or adapt them, treating deviations and adjustments not as exceptions but as essential expressions of expertise.}
\rrsep{}{This distinction marks the difference between a practice of craft that is mechanical and static and one that is dynamic and evolving. As such, existing methods of documentation, like the recipe shown in Figure~\ref{fig:teaser}a, are rendered ineffective for knowledge sharing.}

To better document these workflows, there is a need for frameworks that capture and share \rrsep{both}{this balance of} the structured rules and improvisational deviations embedded in craft practice \rrnov{}{while supporting the evolution of collaborative knowledge archives such as Thingiverse\footnote{https://www.thingiverse.com/},  DeviantArt\footnote{https://www.deviantart.com/}, and Ravelry\footnote{https://www.ravelry.com/}~\cite{yen2024processgallery}}.
The concept of grammars from computational design theory~\cite{stiny1980introduction, knight1995transformations, stiny2006shape} offers a potential path to capture and represent these \rrnov{rules}{rule-based modes} of making.
\rrsep{}{Grammars, as discussed by \citet{stiny1980introduction}, are systems for describing and generating designs using forms such as $A \rightarrow B$ that show how shape $A$ can be \rrnov{replaced by}{transformed into} shape $B$ (replacements can include reflections, rotations, and deletions).}
\rrnov{}{Importantly however, grammars do not produce immutable sequences of action but rather allow for the articulation of departures and variations --- rules but also departures from rules --- that are rule-based modes of practice involve and require.}
\rrnov{Knight describes a \emph{making grammar} as \emph{doing and sensing with stuff to make things}~\cite{knight2015making}, emphasizing the \emph{rules} of material and sensory engagement inherent in the design process.}{For \citet{knight2015making}, \emph{making grammars} --- organized to capture the complex work of \emph{doing and sensing with stuff to make things} --- are systems for describing the structures of material and sensory engagement inherent in the design process. This approach also allows space --- by variations in sequence, timing, and situated application --- for the kinds of improvisational patterns that mark expert craft practice. It also supports a richer responsiveness to contextual and material variations that is essential to craft practice and evolution.}
\rrsep{These grammars not only capture the (skilled) mechanical endeavors to produce an object or thing but also the improvisational patterns and deviations that mark expert craft practice.}{}
\rrnov{Informed by interviews with expert craftspeople (N=16) and existing literature, we develop and demonstrate an \grammar{} to formalize patterns across craft practices that can support the sharing of both rules and improvisational actions (Figure~\ref{fig:teaser}b).}{}

\rrnov{}{In the paper that follows, we draw on interviews with expert craftspeople (N=16) and existing literature to develop and demonstrate an \grammar{} that formalizes patterns across craft practices to support the sharing of both rules and improvisational actions (Figure~\ref{fig:teaser}b).}
Our goal is to \emph{(de)compose craft} practices in ways that support collaboration and learning within extended communities of practice.
\rrsep{In this paper}{In the following sections}, we explore the challenges of documenting \rrsep{improvisation}{tacit, improvisational, and situated practices} in craft workflows.
We begin by reporting on interviews with expert craftspeople across a broad range of \rrsep{practices}{fields} to learn about their documentation methods.
We then \rrsep{develop}{describe} an \grammar{} to \rrsep{formalize}{capture} the tacit knowledge embedded in craft practices.
To demonstrate the grammar in practice, we develop a documentation interface called \emph{\interface{}} powered by a multimodal large language model \rrnov{}{(MLLM)} prompted with our grammar (Figure~\ref{fig:teaser}c).
Finally, we conduct an evaluation study with expert crocheters (N=7) to evaluate how this grammar may be used to \rrsep{spotlight nuanced techniques and tacit knowledge in craft workflows}{highlight their expert practices} for other \rrsep{experts}{knowledgeable craftspeople} to learn from and integrate into their own practices.
\rrnov{}{Our evaluation revealed that the grammar enables adjustable documentation granularity to cater to different documentation consumers, the capture of improvisational and iterative workflows, and knowledge sharing between experts, all while surfacing the inherent limits of externalizing tacit knowledge.}

Through this work, we make three central contributions to CSCW: (1) an \grammar{} for capturing and sharing \rrnov{situated and tacit knowledge in}{}{the tacit knowledge, improvisational actions, and situated adaptations that constitute} craft practices, (2) a demonstration of the grammar's practicality through a documentation interface \rrnov{}{called \emph{\interface{}}}, and (3) insights from \rrsep{expert}{}craftspeople into how this grammar spotlights \rrsep{implicit}{expert} decisions and critical moments of their workflows \rrnov{using}{from} unstructured videos, \rrnov{improving}{supporting} knowledge and expertise sharing within craft communities.
By exploring how to effectively capture and share these skilled practices, this work seeks to inspire and enable wider and more varied practices of collaboration within craft worlds, and the development of computational \rrsep{systems}{tools and infrastructures~\cite{lee2006human, edwards202434, edwards2013knowledge, jackson2007understanding}} that are as flexible and adaptable as the practices they aim to support.

\section{Related Work}
Our work draws on prior research in CSCW, HCI, and the social sciences to explore improvisational interactions.
In this section, we discuss the related literature to situate our work at the intersection of theory and practice by discussing tacit knowledge and improvisation, craft knowledge and expertise sharing, challenges of documentation, and grammars to map improvisation.

\subsection{Tacit Knowledge and Improvisation}
Expert craftspeople, through years of training and practice, cultivate a deep understanding of their craft's materials, tools, and techniques~\cite{risatti2009theory, karana2015material, rosner2012material}.
This understanding transcends the limits of language, 
constituting a type of knowledge that is tacit and \emph{beyond words}: as Michael Polanyi has observed, \emph{``We can know more than we can tell''}~\cite{polanyi1967tacit}.
Recent research has explored how such tacit knowledge is not only embodied by expert craftspeople~\cite{yang2024looking} but also shared in other contexts, highlighting its significance in sustaining cultural heritage~\cite{guo2023tacit} and education~\cite{lensjo2025craftsman}.
This tacit knowledge enables experts to \emph{reflect-in-action} by continuously evaluating their emerging artifacts~\cite{schon1979reflective}, make calculated risks with unpredictable material behavior~\cite{pye1995nature}, and invent techniques to achieve their vision~\cite{li2021we}.
As craftspeople refine their practice, they cultivate a  \emph{professional vision}~\cite{goodwin2015professional} that enables them to perceive, interpret, and collaboratively construct shared understandings of their craft practice~\cite{comi2019constructing}.
This specialized vision allows them to recognize paths invisible to others.

Tacit knowledge empowers craftspeople to apply knowledge from past experiences to new contexts, leading to improvisation within their workflows. 
This makes craft, for \citet{ingold2013making}, an ongoing dialogue between maker, materials, and environment that is fundamentally provisional and emergent in character: \emph{``To practice craft is to work with materials in anticipation of what might emerge, rather than imposing form upon inert matter.''}
Although craftspeople typically begin with a plan, the workflow quickly transforms into a dynamic interplay between plans, perception, and situated actions~\cite{suchman1987plans} where the unpredictable nature of craft requires them to take \emph{risks}~\cite{pye1995nature} by responding to emerging challenges and opportunities~\cite{gaver2022emergence}.
This ability for craftspeople to deviate from plans or expectations~\cite{goveia2022portfolio} is a fundamental aspect of craft expertise~\cite{sennett2008craftsman}.

As discussed in CSCW, this interplay between structure and improvisation results in an evolving practice where knowledge and action are generated and refined~\cite{cheatle2015digital, cheatle2023re}.
Each improvisational response feeds back into the individual craftsperson's tacit understanding, gradually enriching their repertoire of skills and knowledge~\cite{kang2021techarttheory, berliner2009thinking}.
This knowledge is embodied in the continuous cycle of action~\cite{de2018embodied}, breakdown~\cite{jackson2014breakdown}, reflection~\cite{schon1979reflective}, and repair~\cite{jackson2014rethinking} which ensures that craft practices remain vibrant and current in an ever-evolving world~\cite{rosner2012craft, liu2024learning, cheatle2015digital}.
Our approach draws on Polanyi's paradox~\cite{polanyi1967tacit} to (re)imagine documentation artifacts that externalize, without \emph{freezing}\rrnov{,}{} or standardizing, the tacit knowledge that enables craftspeople to respond to unexpected challenges.
External representations~\cite{suthers2003deictic} offer a more realistic, detailed and dynamic way for craftspeople to share their expertise and practice.

\subsection{Craft Knowledge and Expertise Sharing}
Within the CSCW community, knowledge and expertise sharing is a topic of interest through computer-supported documentation, frameworks, and mediums~\cite{ackerman2013sharing, randall1996organisational}.
However, long before the emergence of digital platforms, craftspeople relied on artifacts and demonstrations to share their expertise.
Apprenticeships and guilds, for example, provide hands-on learning experiences where knowledge is passed down through direct observation and imitation~\cite{yarmand2024d, lave1991situated, thoravi2019loki, gasques2021artemis}.

In recent decades, computational tools have been increasingly integrated into the \emph{vocational training} of craft~\cite{zabulis2022digitisation}, approximating hands-on learning experiences~\cite{chai2023design}, supporting expert feedback exchanges~\cite{jane2024feedback, cheng2020critique}, and enabling the asynchronous distribution of craft expertise across geographic distances~\cite{hoffmann2019cyber}.
Among these, video has proven popular for capturing the temporal and embodied nature of craft practices, offering rich visual and auditory channels for documenting complex workflows~\cite{groth2022video, yang2019understanding}.
\rrnov{In hobbyist communities, platforms such as Instructables allow users to share projects through step-by-step guides~\cite{tseng2014product, tseng2015process}}{In creative online communities, where knowledge sharing is central to collaboration, researchers have invested in platforms supporting step-by-step guides~\cite{tseng2014product, tseng2015process} and works-in-progress~\cite{kim2017mosaic}}.
Other approaches designed to approximate embodied learning include manually annotated craft videos with space-time motion paths~\cite{almevik2013tacit} and mixed reality systems for experiential learning~\cite{mueller2003marvel}. 
Complementary efforts focus on structuring craft knowledge for consumption including collaborative platforms for documenting craft ontologies~\cite{partarakis2022web, chaisystematic} and domain-specific search engines for craft-related information~\cite{torrey2009learning}.
However, representing craft practice has often failed to capture its tacit and emergent dimensions, instead reinforcing static notions that miss craft's inherent creativity and dynamism~\cite{corbett1997intelligent, kim2017mosaic}.
Through the formulation of an \grammar{}, we provide a structured representation that spotlights the tacit dimensions of craft, centering often-overlooked aspects within documentation and for knowledge sharing.

\subsection{Challenges in Creating Documentation}
Tacit knowledge is notoriously challenging to document~\cite{borges2002exactitude, makela2018documentation, korzybski1931non}.
As \citet{wood2009tacit} argue, tacit knowledge is difficult to articulate and often remains hidden within the \emph{knowing how} rather than the \emph{knowing that}.
Researchers have explored a range of approaches such as knowledge graphs~\cite{hogan2021knowledge} and process models~\cite{card2018psychology} to document expertise.
These approaches have inspired research across domains applied to documentation artifacts, such as assembly instructions~\cite{agrawala2003designing, ben2021ikea}, graphical document histories~\cite{grossman2010chronicle}, and architectural drawings~\cite{retelny2016embedding}.
Researchers interweave digital and physical artifacts for documentation purposes: embedding in-situ narratives into knitted fabrics~\cite{rosner2009reflections}, enabling asynchronous hardware troubleshooting with 3D scanning~\cite{kwatra2024splatoverflow}, digitizing human-centered CAM workflows~\cite{feng2024cameleon}, and preserving fabrication experimentation using computational notebooks~\cite{tran2024tandem}.

\citet{meiklejohn2024design} explore personal design ledgers to investigate the \rrnov{complexity posed by materials within craft workflows}{``managerial'' tasks that underlie craft workflows}.
\rrnov{While their approach offers valuable insights for capturing tacit knowledge in intelligible forms~\cite{wakkary2021things},}{While their approach offers valuable insights for capturing tacit knowledge in craft project management,} it primarily draws on retrospective accounts that may muddle~\cite{schmidt2012trouble} the practical knowledge of situated decision-making and adjustments that go unnoticed (what \citet{star1999layers} call the \emph{invisible work} of CSCW systems).
\rrnov{}{In contrast, our work focuses on capturing and sharing improvisational actions to improve understanding of the actual craft techniques.}
Building on this premise, we explore patterns in craft workflows not to encode the full breadth and depth of expert knowledge into inflexible structures, but to spotlight the improvisations and decision-making that shape its situated dimensions.
By identifying these patterns in craft workflows with their context, the grammar provides a scalable and extensible method for capturing and sharing expert knowledge.

\subsection{Grammars to Map Expertise}
Our work draws on the concept of grammars from computational theory.
Such grammars are frameworks to represent and generate designs from primitive components~\cite{stiny1980introduction} that allow wide latitude for combinatorial innovation; that is, the ability to be enacted or composed in widely (but not infinitely) varied ways to produce different outcomes.
Thus they provide structure and a language for the expression of action, without being fully deterministic in its outcomes.
Just as the elements of English grammar give structure and comprehensibility to everything from a Shakespearean sonnet to the Apple iTunes terms of service, design grammars provide structure and guidance on the \emph{kinds} of things that exist and how they might be composed, without determining the final shape of the result.
Crucially, while grammars formalize and express \textit{elements} of complex and distributed systems or practices, they do not impose a standardized \emph{script} for action~\cite{he2021finding}, or a fixed rule-set that determines how elements are to be achieved or composed in (all) instances; in Abbott's earlier terms, they support \textit{rule-based}, not \textit{rule-bound}, behaviors~\cite{abbott2014system}.

\rrnov{In design, grammar-based approaches prove useful for developing computational tools that leverage the structure and expressiveness of grammars to generate variations, as demonstrated by tools like PotScript~\cite{mcelroy2023potscript} which use a grammar to create parametric designs for pottery.}{In design-related domains, grammar-based approaches formalize how designs can vary while maintaining a coherence with defined constraints. For example, within architecture, \citet{koning1981language} deconstructed Frank Lloyd Wright's Prairie-style house designs into compositional sequences: beginning with a central fireplace, designs emerge by applying rules such as \emph{extend a wing to create a living room} or \emph{add a terrace parallel to the main axis}, with each rule defining how spatial blocks attach to create the iconic style. Similar grammar-like structures have also been applied to craft where parametric transformations can guide form generation. PotScript~\cite{mcelroy2023potscript} and CoilCAM~\cite{bourgault2023coilcam}, for instance, leverage grammars to specify pottery forms: a cylinder becomes a vase through operations such as tapering, bulging, or adding a rim. Each operation represents a rule that can be applied in different sequences and magnitudes.}
\citet{knight2015making} have developed a computational approach to making using a grammar to bridge the gap between design and fabrication.
Their grammar offers a framework for capturing creative practices by defining making processes as \emph{doing and sensing with stuff to make things}.
They provide a blueprint for learning and replicating these skills by identifying elements and relationships within practices of making.
This may prove valuable in the sharing and evolution of craft practices, where knowledge is transmitted through apprenticeship and experimentation.
In \rrnov{both}{these} instances, grammars help to codify this knowledge, making it more accessible to practitioners distributed in space and time. 
This also aligns with Ingold's\rrnov{~\cite{ingold2009Textility}}{} critique\rrnov{}{~\cite{ingold2009Textility}} of the distinction between design and making precisely because of the role of materials in shaping creative outcomes.

\rrnov{Knight and Stiny's}{The aforementioned design grammars'} emphasis on precise task description lays the foundation for a nuanced understanding of craft processes and facilitates computational approaches to sharing craft knowledge. 
\rrnov{But it provides}{However, they focus primarily on artifact geometries, providing} little account of the reasoning and intuition behind decision making\rrnov{,}{} and the role of intentionality in the transmission of craft knowledge.
As \citet{goodwin2015professional} makes clear, to become a potter is to learn to think and sense like a potter, in relation to the variable material world.
This highlights the need for a more expressive grammar, one that captures and conveys this interplay between structured knowledge and improvisational skill.
\rrnov{Our grammar introduces new representational patterns specifically designed to capture these nuances.}{Our grammar addresses this need by building on \citet{knight2015making}'s grammar to introduce new patterns that formalize how experts navigate uncertainty, deviate from plans, and adjust to emergent conditions during their workflows. By capturing these improvisational aspects of expert practice, our grammar enables experts to make visible and share knowledge that traditional documentation methods struggle to capture.}

\section{Expert Interviews} \label{sec:interviews}
We conducted 13 \rrnov{}{exploratory} interviews with expert craftspeople to \rrnov{survey}{learn about} craft experience across practices, including woodworking, metalworking, and fiber arts.
\rrnov{These interviews focused on understanding how material properties shape their craft workflows.}{Following established approaches for exploratory research in CSCW~\cite{maris2023tech, del2023sound, kang2024challenges}, these interviews were designed to surface unexpected themes and generate initial insights across a breadth of craft materials, tools, and practices before narrowing our focus.}
\rrnov{Drawing from the breadth of}{After} these 13 interviews, we conducted an additional set of \rrsep{in-depth}{focused} interviews to examine an emerging theme in greater detail: craft documentation.
\rrsep{We interviewed three expert craftspeople to explore the documentation norms of their craft practices}{To develop an understanding of the wide range of craft documentation practices, we interviewed three expert craftspeople from distinct domains, each with its own documentation formats and community norms}: baking (recipes), crocheting (stitch patterns), and machining (technical sketches).
These \rrsep{in-depth}{focused} interviews laid a foundational understanding of the challenges in documenting craft, particularly for sharing expertise with others.

\rrnov{The 13 survey and 3 semi-structured interviews were all conducted either in-person or via Zoom and ranged in length from 30 to 90 minutes.}{The 13 exploratory and 3 focused interviewers were all semi-structured and conducted either in-person or via Zoom. Each interview ranged in length from 30 to 90 minutes.}
The \rrnov{survey}{exploratory} interviews (S1-S13) followed \rrnov{an}{the} interview \rrnov{structure}{protocol} included in Supplementary A1.
The \rrnov{in-depth}{focused} interviews (P1-P3) explore documentation practices in two phases of craft: (1) During fabrication to focus on tools and motivations for documenting and (2) Post-fabrication to examine how craftspeople revisit and share their documentation with others (interview \rrnov{structure}{protocol} in Supplementary A2). 
\rrsep{Participants were sampled from our personal networks and expanded through snowball sampling~\cite{goodman1961snowball}.}{To sample participants from the existing structured and relational networks of expert craftspeople, we used snowball sampling starting with initial respondents from our personal \rrnov{relationships}{networks} and prior research engagements in craft communities \rrnov{}{[reference removed for anonymity]}. For example, S7, within the first author's network, introduced us to S4, a previous collaborator \rrnov{}{of S7}. This method follows well-established principles of ethnographic fieldwork in communities of practice, and reflects the linked and networked character of the craft worlds studied~\cite{becker1982art, lave2019apprenticeship}.}
\emph{Experts} were defined as craftspeople with a minimum of five years of professional experience within their craft.
Table~\ref{tab:interview_participants} summarizes the backgrounds and craft techniques of all participants.


\begin{table*}[t]
\centering
\begin{tabular}{
>{\raggedright\arraybackslash}
p{.03\columnwidth}
>{\raggedright\arraybackslash}p{.04\columnwidth}
>{\raggedright\arraybackslash}p{.29\columnwidth}
>{\raggedright\arraybackslash}p{.35\columnwidth}
>{\raggedright\arraybackslash}p{.18\columnwidth}}
\toprule
\textbf{ID} & \textbf{Exp. (Yrs)} & \textbf{Profession} & \textbf{Craft Techniques} & \textbf{Documentation Tools} \\
\midrule
S1 & 50+ & Sculpture and Fiber Artist, Teacher & Knitting, Crocheting & - \\
S2 & 20 & Product Designer, Architect, Artist, Entrepreneur & 3D Printing, Metalworking, Sewing & Paper and Pencil \\
S3 & 15 & Product Designer & 3D Printing, Sewing, Woodworking & Photo, Video \\
S4 & 15+ & Sculpture Artist, Professor & Hand Knitting, Crocheting, CNC & Paper, Photo \\
S5 & 15+ & Furniture Designer & CNC, Laser Cutting, 3D printing, Weaving, Woodworking & Photo \\
S6 & 10+ & Furniture Designer, Professor & 3D printing, CNC, Laser Cutting, Woodworking & Paper, Photo \\
S7 & 25+ & Sculpture Artist & Welding, Laser Cutting, Sewing, Sketching & Photo \\
S8 & 20+ & Knitter, Knitting Studio Manager & Knitting, Crocheting & - \\
S9 & 15 & Jewelry Designer, Professor & Enameling, 3D Printing & Paper, Photo\\
S10 & 20 & Sculpture Artist, Professor & Laser Cutting, CNC,  Sketching & Photo\\
S11 & 50+ & Textile Designer & Loom Weaving, Machine Knitting & Photo\\
S12 & 25 & Metalworker, Makerspace Director & Welding, Metal Fabrication, CAD & Photo, Video\\
S13 & 20 & Weaving Artist, Studio Manager & Floor Loom Weaving & - \\
\midrule
P1 & 15 & Machinist, Entrepreneur & Metal Machining and Welding & Paper, Photo, Video\\
P2 & 15 & Baker, Graduate Student & Baking & Photo, Snapchat\rrsep{}{, Instagram} \\
P3 & 5 & Crocheter, Software Engineer & Crocheting, Knitting &  Photo, Video, Tablet\rrsep{}{, Instagram}\\
\bottomrule
\end{tabular}

\caption{Summary of expert \rrnov{survey}{exploratory} (S) and \rrsep{}{focused (P)} interview study \rrsep{(P)}{}participants and the craft techniques and documentation tools discussed.}
\label{tab:interview_participants}
\end{table*}

\subsection{Findings}

\rrsep{To identify key patterns of documenting craft, the first author analyzed the interviews following the principles of grounded theory~\cite{glaser1968discovery}, resulting in 198 inductive codes that revealed four themes as shown in Table~\ref{tab:codes}: Improvisation and adaptation, tools for sharing knowledge, limitations of documentation, and sharing knowledge for learning.}{To identify patterns of documenting craft, the first author analyzed the focused interviews following the principles of grounded theory~\cite{glaser1968discovery}, performing initial open coding \rrnov{and continuing to sample new participants until thematic saturation was reached}{on the three interviews}. This resulted in 198 codes under four themes (Table~\ref{tab:codes}): Improvisation and adaptation, tools for documenting practice, limits of documentation, and sharing knowledge for \rrnov{}{community} learning.} \rrnov{}{The exploratory interviews provided contextual understanding across craft domains and supportive examples for themes identified in the focused interview analysis.}

\begin{table*}[t]
    \centering
    \begin{tabular}{
    >{\raggedright\arraybackslash}p{0.2\columnwidth}
    >{\raggedright\arraybackslash}p{0.2\columnwidth}
    >{\raggedright\arraybackslash}p{0.2\columnwidth}
    >{\raggedright\arraybackslash}p{0.3\columnwidth}}
    \toprule
    \textbf{Theme} & \textbf{Description of Theme} & \textbf{Code Groups} & \textbf{Examples of Low-Level Codes} \\
    \midrule
    Improvisation and Adaptation & Actions taken in response to unexpected situations or material variations & \emph{Communicating Uncertainty}, \emph{Adapting the Fundamentals}\rrsep{, \emph{Tools for Sharing Knowledge}}{} & \emph{some of the movements are almost invisible for machining} (P1), \emph{change templates to use the scraps} (P1) \\
    \rrsep{Tools for Sharing Knowledge}{Tools for Documenting Practice}	& Methods to document craft practices & - & \emph{pictures/videos supplement notation due to variability} (P3), \emph{binder has all recipes with sticky notes as dividers} (P2) \\
    \rrsep{Limitations}{Limits} of Documentation & Challenges experts face when documenting craft & \emph{Level of Detail}, \emph{Recording While Crafting}, \emph{Embodied Knowledge} & \emph{takes 2-3 times longer to make videos} (P2), \emph{people are buying static patterns which is still annoying in case there are any mistakes} (P3) \\
    Sharing Knowledge for \rrnov{}{Community} Learning	& How documentation supports \rrnov{non-expert}{}learning \rrnov{}{within a community} & \emph{Supporting \rrnov{Pedagogy}{Teaching and Learning}}, \emph{\rrsep{Lasting Benefits of Documentation}{Encouraging Long-Term Thinking}} & \emph{free YouTube tutorials to learn things that she doesn't know} (P3), \emph{lot of knowledge to impart and there is so many different things} (P1) \\
    \bottomrule
    \end{tabular}
        
    \Description{}
    \caption{The four high-level themes from our interviews that motivated our \grammar{}, alongside a theme description, code groups, and exemplary low-level codes.}
    \label{tab:codes}
    \end{table*}

\subsubsection{Improvisation and Adaptation} 
\label{interviews:improv}

We observed two key factors when it comes to the theme of improvisation and adaptation: how to \rrsep{document}{communicate} uncertainty and how to adapt from the fundamentals for unanticipated situations.

\paragraph{Communicating Uncertainty.}
Half of the participants (8) mentioned the improvisational nature of their craft by describing how they navigate uncertainty when adapting to emerging material behaviors, tools and techniques, and design constraints.
For instance, P2 (Baker) explained, \emph{``If there's a recipe that I didn't really like [and] I wanted to adapt [it], I usually just write on it... with a pen... `70 grams of sugar instead of 90'... [but] I didn't want to write it on the [recipe] just in case it didn't work.''}
This reluctance to alter the recipe highlights the difficulty of documenting improvisation, especially given the possibility of unsuccessful outcomes.

\paragraph{Adapting the Fundamentals.}
S6 (Furniture Designer) likened craft practices to the improvisational aspects of jazz music where craftspeople \emph{``learn the vocabulary [of their craft]: Materials, processes, tools to be lyrical and really unburdened by [anything].''}
This sentiment was shared by S5 (Furniture Designer), who emphasized that shifting away from rigid structures and learning about material behaviors is crucial to navigate the unpredictable landscape of craft---\emph{``It's about forgetting your product design education.... It's not until you end up with something like this [standard material] that you start thinking like a designer.''}
To design for this uncertainty, S9 (Jewelry Designer) starts with a partial design: \emph{``I'm playing around with the material... [with] between 50 to 70 percent of an image in my head.... And with every new hammer that you get, you find your work also changes because there [are] all these new possibilities with every tool that you didn't know about your material.''}

\subsubsection{\rrsep{Tools for Sharing Knowledge}{Tools for Documenting Practice}.}
\label{interviews:documentation}
Craftspeople employ a variety of documentation \rrsep{tools}{artifacts} to capture their workflows, such as photos (12), sketches and text (7), and videos (4).
Photos offer a quick way to document specific steps, highlight important measurements, or record precise configurations.
As P1 (Machinist) explained, \emph{``A photo can give me enough information [for remembering] where I'm putting what dies for bending''} which ensures accuracy when referring back to the documentation.
Written notes and sketches remain valuable in craft practices for quickly visualizing ideas, particularly for designing and exploring.
S2 (Product Designer), explained that sketching is essential for his practice: \emph{``I have to sketch; I can't work things out in my head.''}
Finally, videos provide a more dynamic and comprehensive record, especially for actions difficult to describe in static forms of documentation, such as complex movements or subtle material interactions.
P3 (Crocheter) noted that video \emph{``fills in the gaps of what text is unable to describe''} by capturing nuanced techniques and providing a visual record for the entire workflow, often through time-lapse documentation.
\rrsep{}{Participants (9) described sharing these documentation artifacts through social media platforms such as Instagram, Snapchat, and Discord.}

\subsubsection{\rrsep{Limitations}{Limits} of Documentation} 
\label{interviews:limitations}
Participants brought up \rrsep{limitations}{the limits} of documentation in three categories: capturing the right level of detail, having to think about knowledge sharing during the craft workflow, and the constraints on flexibility different documentation tools provide.

\paragraph{Level of Detail.}
Eight participants noted the challenge of fully capturing workflows and values embedded in craft practices. 
P1 (Machinist) emphasized the importance of recording details such as the amount of force applied when bending metal, stating, \emph{``Some of the details that maybe you can't see from an image or a video... I applied a decent amount of force here. I applied maybe a little less force here.''}
S1 (Sculpture and Fiber Artist) noted cultural assumptions implicit in documentation; a Japanese craft book she uses does not include sizes, while a Scandinavian book had no descriptions of the specific techniques, presumably assuming these are known to readers.
This variation underscores the challenge of creating universally accessible documentation.

Even with readily available resources such as online tutorials, \rrsep{limitations}{limits} persist, as S4 (Sculpture Artist) shared how he had to attend in-person workshops to learn to knit: \emph{``I’m not gonna be able to do this just by watching YouTube....It's just a `Hey, show up and we'll teach you how to knit.' So I sat there for an hour or two and someone would like occasionally come by and say like, `Oh, how's it going? Oh, you're doing this slightly wrong....' I went home and I just practiced a bunch.''} 
These challenges collectively emphasize the need for more comprehensive and nuanced documentation methods that can effectively capture craft knowledge across cultures and skill levels to avoid ambiguity.

\paragraph{Recording While Crafting.}
Five participants noted that while high-fidelity documentation is valuable, creating such artifacts can be challenging. 
P2 (Baker) described the significant overhead involved in filming detailed videos: \emph{``Filming something usually makes something take two to three times longer... you're having to wash your hands all the time because your hands are dirty, but then you're like, `Is my camera dead.' As a non-expert, I'm like, `Is my camera focused' ... taking videos of myself is really hard work versus pictures.''}
Similarly, S8 (Knitter) added that it is \emph{``hard to watch a video and follow it along... especially when your finger is like blocking the frame, I think there is a level of dexterity there that just... [is] like muscle memory,''} further emphasizing the difficulty in capturing subtle hand movements in videos. 
This highlights the tension between the desire for comprehensive documentation and the practical constraints of maintaining workflow and minimizing disruption.
P3 (Crocheter) also discussed the post-processing of this video data as her edited videos \emph{show the yarn and then just the finished product [with] things in between.... I'll show little [clips] and then speed it up... [or] full time lapse videos... for a smaller item because... the amount of footage I have would be too much to deal with.''}
Beyond the time investment, creating effective videos also requires technical skill and familiarity with video taking and editing --- being a craft in and of itself.

\paragraph{Embodied Knowledge.}
Three participants demonstrated their preference for relying on memory and tacit knowledge rather than explicit documentation, particularly when teaching.
S1 (Sculpture Artist) explained, \emph{``I do not write down my patterns... I will have [students] stand next to me while I am doing all of the math,''}.
S8 (Knitter) learned to knit through YouTube --- \emph{``a great place to learn. And then I would start to challenge myself for slightly different, more advanced projects....''} --- and now improvises while teaching non-experts how to knit \emph{``[her] version of other people's patterns''}
These perspectives demonstrate the \rrsep{limitations}{limits} of documentation tools in capturing the nuances of tacit knowledge and improvisation inherent in expert craft workflows.

\subsubsection{Sharing Knowledge for \rrnov{}{Community} Learning}
\label{interviews:learning}
We break \rrnov{this }{}down this final theme into observations surrounding \rrnov{pedagogy}{how documentation supports teaching and learning} and the \rrnov{lasting benefits of such documentation}{its role in encouraging long-term thinking}.

\paragraph{Supporting \rrnov{Pedagogy}{Teaching and Learning}.} Eight participants shared how documentation plays a crucial role in supporting the learning and knowledge sharing within craft practices. 
P3 (Crocheter) emphasized the importance of creating clear patterns: \emph{``[they need] to be easy to make. So [the design] can't just like look good; it also needs to be the best way of doing it.''}
This highlights the value of documentation that not only captures the steps involved, but also considers the non-experts' perspective to ease the sharing of knowledge.
P3 developed personal preferences by consuming documentation: \emph{``It first started with me like not liking specific things of other people's patterns.... When I wanted to make multiple of something, I would just kind of rewrite [their pattern] or change it a lot\rrnov{.}{,}''} \rrnov{}{reflecting how even experienced practitioners engage as learners when encountering others' approaches.}
S6 emphasized the value of bridging familiar concepts with new techniques in craft documentation, suggesting that grounding unfamiliar tools and processes with familiar materials and metaphors makes them more accessible: \emph{``It's a little daunting to approach something or approach a tool you don't know.... [but] you could do a lot with paper to understand an object.''}
These insights underscore the critical role of documentation in craft as a pedagogical tool that empowers through clear communication.

\paragraph{\rrsep{Lasting Benefits of Documentation}{Encouraging Long-Term Thinking}.} S5 (Furniture Designer) discussed the importance of documenting the use of natural materials to encourage long-term thinking: \emph{``It's about getting people to think a lot, much more long term: `What am I doing now?' and `What's going to be the impact of these actions when I'm gone?'''}
This highlights the potential for documentation to transmit not only technical skills, but also values and cultural perspectives.
S12 (Metalworker) shared that \emph{``\rrnov{The}{the} thing that is really rewarding as a teacher and sharing knowledge around stuff is that it's just so damn empowering,''} capturing \rrnov{the impact of sharing craft knowledge on teacher and student}{how knowledge sharing within craft communities benefits both creator and consumer}.


\section{Elementary Grammar} \label{sec:grammar}

\rrnov{The resulting \grammar{} includes the following seven key patterns: \detailblocks{}, \reflective{}, \note{}, \external{}, \segments{}, \branches{}, and \revisionloops{}.}{From our findings, we developed an \grammar{} to capture and share expert knowledge in craft workflows. Table~\ref{tab:grammar-mapping} maps each of the seven grammar patterns (\detailblocks{}, \reflective{}, \note{}, \external{}, \segments{}, \branches{}, and \revisionloops{}) to the interview findings that motivated its design.}

\begin{figure}[h!]
    \centering
    \includegraphics[width=\linewidth]{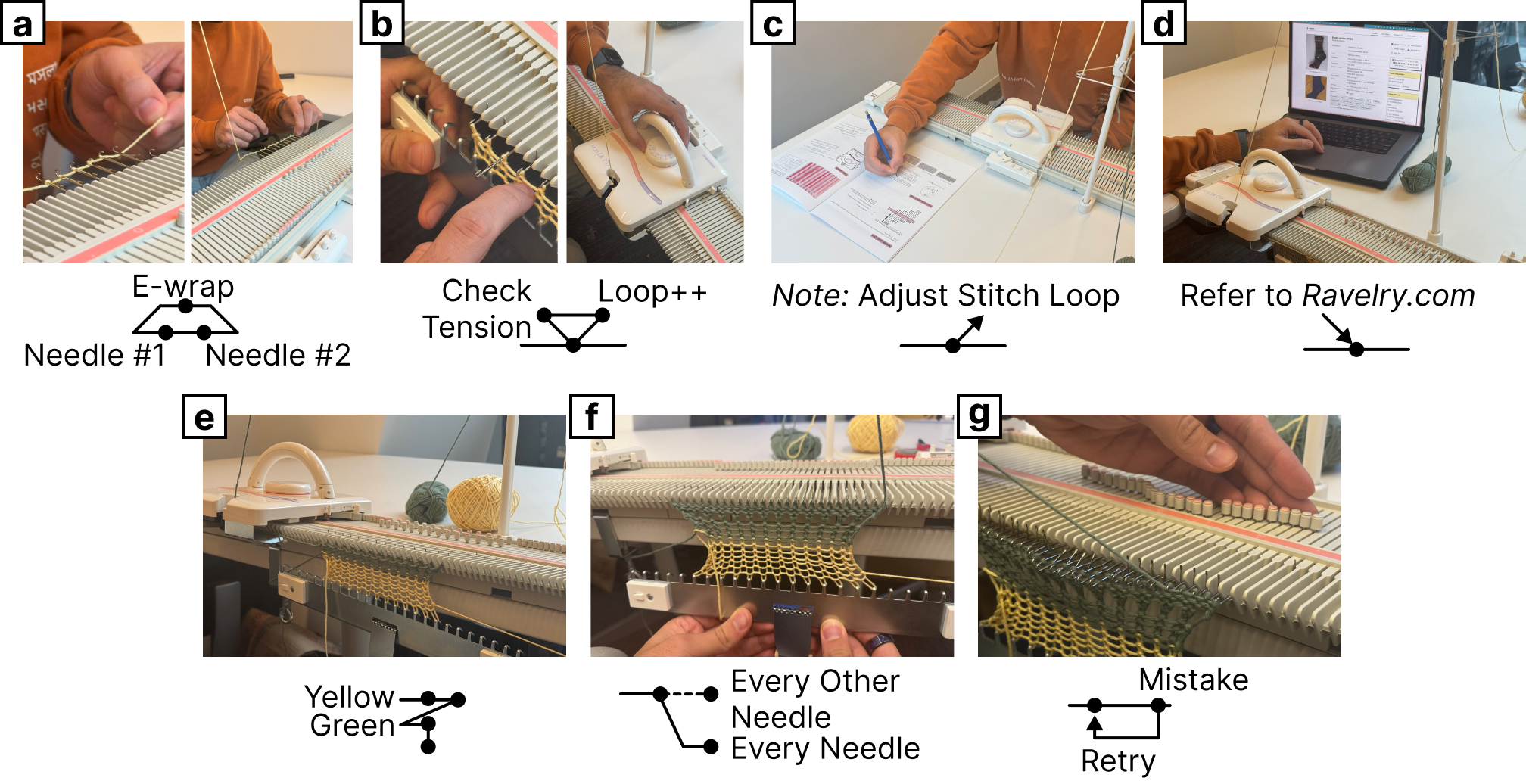}
    \caption{During this machine knitting workflow, steps involving tacit knowledge are challenging to document. Our \grammar{} captures this with these patterns: (a)~\detailblocks{}, (b)~\reflective{}, (c)~\note{}, (d)~\external{}, (e)~\segments{}, (f)~\branches{}, and (g)~\revisionloops{}.}
    \label{fig:grammar}
    \Description{}
\end{figure}

\rrnov{}{This grammar builds on Knight and Stiny's making grammar~\cite{knight2015making} by capturing the non-linear relationships and improvisational actions between their primitives (stuff, things, doing, sensing) that characterize expert craft practice. This enables our grammar to not just capture the \emph{what} of making, but document and share the \emph{how} of expert practices.}
Figure~\ref{fig:grammar} highlights these patterns with an example of a machine-knitting workflow.

\begin{table*}[t]
\centering
\begin{tabular}{
>{\raggedright\arraybackslash}p{.15\columnwidth}
>{\raggedright\arraybackslash}p{.25\columnwidth}
>{\raggedright\arraybackslash}p{.25\columnwidth}
>{\raggedright\arraybackslash}p{.25\columnwidth}
}
\toprule
\textbf{Grammar Pattern} & \textbf{Theme} & \textbf{Interview Finding} & \textbf{Addresses Finding By...} \\
\midrule
\detailblocks{} & Limits of Documentation & Level of Detail & Enabling detail personalization \\
\reflective{} & Improvisation and Adaptation & Adapting the Fundamentals & Capturing materials- and tools-driven adjustments \\
\note{} & Limits of Documentation & Level of Detail, Embodied Knowledge & Annotating and describing often-missed details \\
\external{} & Sharing Knowledge for Community Learning & Supporting Teaching and Learning & Linking to external resources \\
\segments{} & Tools for Documenting Practice, Sharing Knowledge for Community Learning & Recording While Crafting & Structuring workflows into sections \\
\branches{} & Improvisation and Adaptation, Sharing Knowledge for Community Learning & Adapting the Fundamentals, Encouraging Long-Term Thinking & Documenting expert deviations and practice evolution \\
\revisionloops{} & Improvisation and Adaptation & Communicating Uncertainty & Preserving experimentation paths \\
\bottomrule
\end{tabular}
\caption{\rrnov{}{Mapping of interview findings to our grammar patterns.}}
\label{tab:grammar-mapping}
\end{table*}

\subsection{\detailblocks{}}

\begin{quote}
\emph{
``[With] a published [craft] book... from Scandinavia, the presumption is everybody knows the techniques.... So the instructions don't have to be exact.... Japanese sewing patterns don't come in sizes. They just have little drawings of what the pattern pieces should look like.\rrnov{...}{}''} --- S1 (Sculpture and Fiber Artist)
\end{quote}

\rrnov{S1 highlights the need for adjustable lenses in craft documentation to cater to different audiences.}{S1 highlights how craft documentation is created with embedded assumptions about what the audience already knows. This underscores a broader challenge of determining the appropriate level of detail for diverse audiences of varying cultural norms and expertise levels.}
Rather than documenting craft knowledge as a monolith of equally-important steps, the grammar assigns a level of granularity to each state in the workflow using \detailblocks{}.
This level of granularity, listed as low, medium, and high, dictates the level of documented detail (Section~\ref{interviews:limitations}).
A high level of granularity might include detailed descriptions of tool usage, material handling, and subtle nuances in movement, as well as explanations of the rationale behind specific choices and adjustments.
Conversely, a low level of granularity might simply state the action without further elaboration, suitable for experts who already possess the necessary background knowledge.

In Figure~\ref{fig:grammar}a, the knitter demonstrates \detailblocks{} by providing a close-up capture of their hands performing the \emph{e-wrap} cast on, accompanied by a precise verbal description of each movement.
Within this detailed documentation, nuanced movements are highlighted as high level of granularity --- for instance, the exact placement of the yarn on the specific needle or the hand motion used to create the wrap.
This allows other craftspeople to tailor the level of granularity in the framework to their needs, ensuring they grasp the nuances for successful execution of the technique.
Experienced craftspeople recognize this as a familiar cast-on technique and may not require such granularity.
By incorporating \detailblocks{}, our grammar enables the creation of documentation that is both comprehensive and adaptable to the needs of diverse learners.

\subsection{\reflective{}}
\begin{quote}
    \emph{``I'm playing around with the material... [with] between 50 to 70 percent of an image in my head.... And with every new hammer that you get, you find your work also changes because there [are] all these new possibilities with every tool that you didn't know about your material.''} --- \rrnov{S12 (Metalworker)}{S9 (Jewelry Designer)}
\end{quote}

As \rrnov{S12}{S9} describes, craftwork often involves a dialogue with materials.
Craftspeople do not always have a fully formed plan; they engage in cycles of action, observation, and adjustment, responding to the nature of their materials and tools.
This aligns with the concept of a \emph{``reflective practitioner''}~\cite{schon1979reflective} who continuously evaluates their actions in light of emerging challenges and adjusts their approach accordingly.
Ingold~\cite{ingold2013making} emphasizes the agency of materials in shaping craft workflows, highlighting the need for craftspeople to respond to the inherent variability and unpredictability of their chosen medium.
Therefore, capturing these loops between the craftsperson and material, where perception and action inform each other, is crucial for conveying the dynamic and responsive nature of craft.

\reflective{} represent the iterative process of experimentation and adjustment that characterizes craftwork.
With \reflective{}, the grammar captures the non-linear nature of craft, where decisions are made in response to emerging challenges and unexpected discoveries.
Moreover, by studying these loops, other craftspeople gain insights into the tacit knowledge that guides decision-making, developing a deeper understanding of how to approach unexpected situations.

In Figure~\ref{fig:grammar}b, after the knitter knitted the first few rows, they pull out a needle to assess the tension. 
They may adjust the tension dial on the knitting machine, demonstrating a dynamic interaction between doing and sensing.
This moment of reflection highlights the knitter's responsiveness to the material and their ability to adjustment.
By incorporating \reflective{} into our grammar, we create documentation that accurately reflects the iterative and adaptive nature of craft practices.

\subsection{\note{}}
\begin{quote}
    \emph{``Some of the details that maybe you can't see from an image or a video... I applied a decent amount of force here. I applied maybe a little less force here.''} --- P1 (Machinist)
\end{quote}

P1 highlights that some actions within craft workflows involve details that may not be readily apparent through observation alone.
These nuances, such as the precise amount of force applied or the specific tool settings used, can significantly impact the result of the action.
For experts to capture these details effectively, our grammar needs to include annotations that provide further detail beyond the grammar's foundational elements and context of the \emph{why} behind every action.

\note{} is a flexible structure for craftspeople to add information and context that does not fit into the grammar's defined patterns and elements.
The foundational elements benefit from additional annotation, particularly for sensing and doing where information from visual demonstrations remains incomplete.
While some aspects of craft may remain elusive in documentation, situated annotations provide a freeform way to embed crucial context.

To illustrate this pattern, Figure~\ref{fig:grammar}c shows how a knitter adds notes about their workflow as they adjust the tension dial on their knitting machine.
In this scenario, the knitter checks, or \emph{senses} the tension of the knit fabric.
They add notes about how different yarns require varying levels of tension due to their specific fiber content, ply, and thickness.
By identifying \note{}, our grammar empowers craftspeople to articulate and share the subtle, yet crucial, aspects of their practice that often escape traditional documentation methods.

\subsection{\external{}}
\begin{quote}
    \emph{``I’m not gonna be able to do this just by watching YouTube....It's just a `Hey, show up and we'll teach you how to knit.' So I sat there for an hour or two and someone would like occasionally come by and say like, `Oh, how's it going? Oh, you're doing this slightly wrong....' I went home and I just practiced a bunch.''} 
    --- S4 (Sculpture Artist)
\end{quote}

As S4 describes, learning is an ongoing journey fueled by interactions with fellow craftspeople through external resources.
Online resources, workshops, and mentorship provide valuable paths for acquiring new skills and expanding one's understanding.

This exchange of individual knowledge and communal resources is essential to the evolution of craft, motivating a need for  \external{}.
This allows our grammar to connect with external sources of knowledge --- websites, videos, online forums, and other repositories of craft knowledge.
By integrating references, the grammar acknowledges that craft practice often involves drawing on a network of information that extends beyond a single document.
This interconnectedness fosters a richer understanding of the craft and encourages continuous learning. 

Figure~\ref{fig:grammar}d provides a concrete example.
The knitter, facing a challenge with changing colors on their specific machine, turns to online resources for guidance.
They engage in online community forums such as \textit{Ravelry}~\cite{ravelry} to find the solution as well as other knitters who have faced this issue.
By incorporating these external resources (e.g., links, videos, or forum discussions) into their documentation, they create a more comprehensive record of their workflow and contribute to the collective knowledge of the craft community.

\subsection{\segments{}}
\begin{quote}
    \emph{``[Videos] show the yarn and then just the finished product [with] things in between.... I'll show little [clips] and then speed it up... [or] full time lapse videos... for a smaller item because... the amount of footage I have would be too much to deal with.''} --- P3 (Crocheter)
\end{quote}

As P3 noted, observing every stitch in a multi-hour knitting tutorial can quickly lead to disengagement.
A balance between comprehensiveness and conciseness is essential, allowing for both in-depth study and efficient comprehension.

\segments{} are a structural pattern that enables the division of a craft workflow into distinct sections, mirroring the creation of individual components within the final artifact (e.g., the sleeves of a sweater).
By selectively expanding and querying segments, users can navigate the documentation artifact more effectively, either focusing on specific techniques or gaining a broader overview of the captured workflow.

For instance, Figure~\ref{fig:grammar}e illustrates the knitter introducing a new yarn type, marking a distinct segment in the workflow.
This segment might contain specific techniques for joining the yarn or adjusting tension.
As this is an optional and more advanced technique, an expert knitter can quickly navigate to this section for reference, while a less experienced knitter might skip it to focus on the overall construction.
This layered approach ensures the documentation remains accessible and relevant for learners at all levels, catering to diverse learning needs and interests.

\subsection{\branches{}}
\begin{quote}
    \emph{``[YouTube is] a great place to learn. And then I would start to challenge myself for slightly different, more advanced projects.... [Now, I teach my] version of other people's patterns''} --- S8 (Knitter)
\end{quote}

As S8's quote illustrates, craftspeople frequently deviate from established patterns, adapting their techniques, materials, and tools to achieve specific goals or explore new creative directions.
These deviations highlight the dynamic interplay between following instructions and applying tacit knowledge.
Capturing these moments is crucial for understanding how craftspeople personalize their craft and contribute to its evolution.

\branches{} document these deviations to capture insights into the tacit knowledge and decision-making that underpin expert practice.
This pattern allows us to represent the non-linearity of real-world craft workflows, acknowledging the alternative paths that craftspeople often explore.
By comparing the original pattern with the actual workflow, the grammar can identify and document specific points of divergence, including variations in materials, actions, tools, or techniques.

In Figure~\ref{fig:grammar}f, the knitter modifies the pattern by knitting into every needle instead of every other needle to achieve a tighter knit.
This illustrates how craftspeople adapt instructions based on their individual goals and material understanding.
By documenting this change using \branches{}, the grammar captures not only the technical modification, but also the underlying intention and creative decision-making process.

\subsection{\revisionloops{}}
\begin{quote}
    \emph{``It's a little daunting to approach something or approach a tool you don't know.... [but] you could do a lot with paper to understand an object.''} --- S6 (Furniture Designer)
\end{quote}

This reflection from S6 highlights the importance of iterative experimentation in craft.
The ability to explore different approaches and \emph{undo} actions is crucial for both learning and mastering a craft.
This experimentation often involves utilizing readily available materials and tools to prototype and refine ideas before translating them to the final artifact.
Such actions significantly influence the workflow and contribute to the situated knowledge of the craftsperson.

\revisionloops{} document two critical aspects of iteration in craft workflows.
First, they capture the prototyping phases where craftspeople experiment with readily available materials and tools to refine ideas before translating them to the final artifact.
Second, this pattern documents instances where craftspeople sense something is off with their workflow, undo recent steps, and redo them differently.
This \emph{undo-redo} cycle preserves the \emph{``paths-not-taken''}~\cite{goveia2022portfolio} and valuable learning experiences often lost in traditional documentation.
By observing these experiments, other craftspeople gain insights into the decision-making process and develop a deeper understanding of the craft.

In Figure~\ref{fig:grammar}g, the knitter, realizing a design choice doesn't align with their vision, unravels a few rows to revert to a previous state.
This action highlights the importance of \emph{repair}~\cite{jackson2014rethinking} within craft practices, extending beyond knitting to woodworking, ceramics, and other domains. 
Documenting these revisions reveals the expert's ability to recognize problems, adapt to unexpected outcomes, and recover from setbacks which are all qualities of expert practice that traditional linear documentation methods struggle to effectively map and communicate.\\

Our \grammar{} goes beyond individual patterns and elements to create a coherent framework for capturing the multi-dimensional nature of craft expertise.
When applied to a complete workflow, as demonstrated in the additional examples in Table~\ref{tab:example_grammars}, these patterns interact dynamically and reveal the richness and complexity of craft expertise.

\begin{table*}[t]
\centering
\begin{tabular}{
>{\raggedright\arraybackslash}p{.17\columnwidth}
>{\raggedright\arraybackslash}p{.25\columnwidth} 
>{\raggedright\arraybackslash}p{.25\columnwidth}
>{\raggedright\arraybackslash}p{.23\columnwidth}}
\hline
 & \textbf{Carving a Wooden Spoon} & \textbf{Folding a Paper Crane} & \textbf{Sketching a Person} \\
\hline
\textbf{\detailblocks{}} & Detailed marks during initial carving & Precise angles for each fold & Coarse layout of proportions \\ \\
\textbf{\reflective{}} & Adjusting to wood grain changes & Adjusting to paper thickness & Adjusting to pencil hardness \\ \\
\textbf{\note{}} & Tool pressure & Measurements for folding & Light source and intensity \\ \\
\textbf{\external{}} & Scandinavian carving videos & Origami pattern books & Anatomy diagrams \\ \\
\textbf{\segments{}} & Rough shaping, detail work, finishing & Base, shaping, final folds & Sketch, detail, shading \\ \\
\textbf{\branches{}} & Deviations around wood knots & Alternate folding method for stability & Testing alternative hand poses \\ \\
\textbf{\revisionloops{}} & Corrections of inaccurate marks & Unfolding due to misaligned wings & Erasing to redraw proportional fingers \\
\hline
\end{tabular}
\caption{Examples of how each \grammar{} pattern appears across different craft practices.}
\label{tab:example_grammars}
\end{table*}

\section{Grammar in Practice: \emph{\interface{}}}

To demonstrate the practical utility of our grammar we designed an interface called \emph{\interface{}} for documentation and collaboration that transforms unstructured craft videos into \rrnov{graph-based visual representations}{graphs}, highlighting the patterns and foundational elements identified in our \grammar{}.
Collaborators (other craftspeople) use these graphs to learn from the captured situated tacit knowledge.
This section describes the design of the authoring interface that enables craftspeople to interact with these graphs for others to consume.

\begin{figure}[h]
  \centering
  \includegraphics[width=\linewidth]{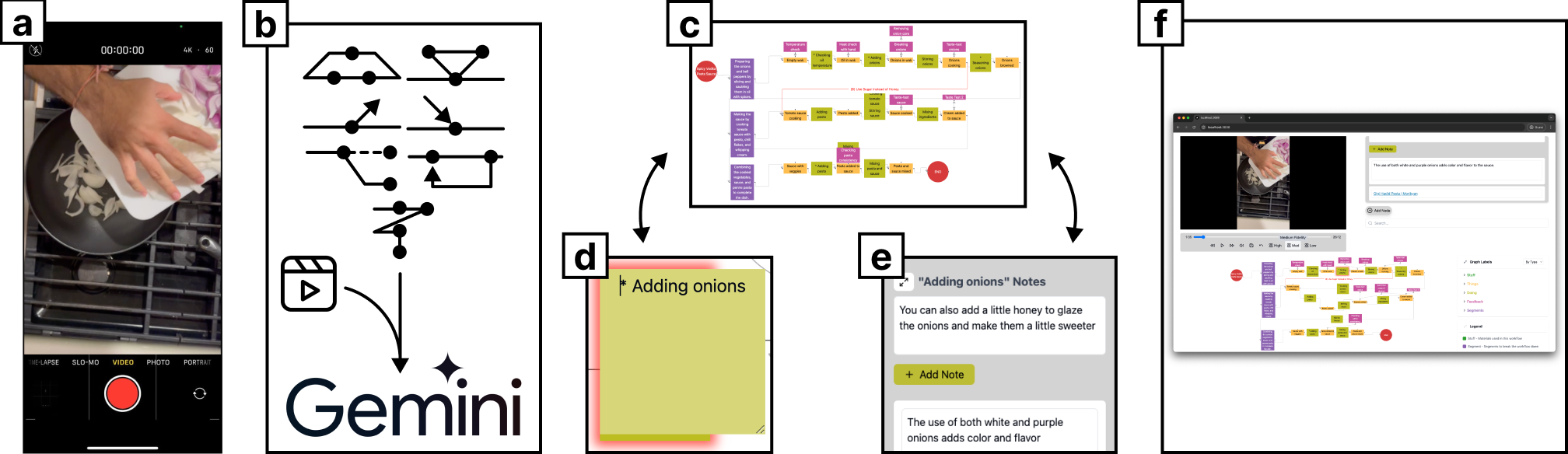}
  \caption{The pipeline of our demonstrative implementation of the grammar.}
  \label{fig:workflow}
  \Description{}
\end{figure}

The pipeline for \emph{\interface{}} comprises five stages, as shown in Figure~\ref{fig:workflow}: (a)~experts record narrated videos of their craft workflows, (b)~these videos are processed into structured data (JSON) by a multi-modal large language model (MLLM) using our grammar as system prompt, (c)~the JSONs are converted into structured graphs, (d,e)~experts refine the graphs through our authoring interface, so (f)~other craftspeople can navigate the revised graphs with a context-rich view of the workflow.

\subsection{Video Capture and Narration} \label{subsec:video}

The documentation interface requires experts to record themselves engaging in their craft, while simultaneously narrating their actions, thoughts, and decisions.
The emphasis is on producing a rich verbal account that complements the visual demonstration.

Our interview findings suggest that self-recording is challenging, therefore our interface allows camera setups to be either static or dynamic, and the camera itself can be used as a pointing device to highlight specific features or materials, allowing the process to integrate naturally into existing workflows. 
Figure~\ref{fig:workflow}a illustrates this approach: a chef prepares pasta while narrating, here adding onions to the sauce.
After recording, the video is uploaded to Google Cloud Storage~\cite{bisong2019overview}.

\subsection{Analyzing the Video} \label{subsec:analyze}

To analyze the unedited video, we use Google's MLLM, Gemini 1.5 pro~\cite{team2024gemini}, guided by a structured prompt based on our grammar (see Supplementary A3).
This prompt defines each foundational element and pattern in our grammar, providing detailed descriptions and illustrative examples to enhance interpretation~\cite{zhou2025instructpipe}.
This process results in a structured representation of the craft workflow, capturing the flow of decision-making and material engagement (Figure~\ref{fig:workflow}b).

To ensure high-quality graph generation, we \rrsep{include a set of}{added} constraints for the MLLM within our prompt.
\rrsep{The graph must be fully connected, meaning all nodes are linked, with a path from the first to the last thing with no isolated elements. It must also be complete, spanning the full video duration, with all the nodes timestamped from the beginning (0 sec) to the end of the video. The model then processes both the uploaded video and the grammar-based prompt to generate a structured JSON output, which is stored in cloud storage.}{The graph must be fully connected, meaning there is a path linking all the nodes. It must also be complete, with node timestamps spanning the full video duration. The model then processes the uploaded video and prompt to generate and store a structured JSON output (see example in Supplementary A4).}
\rrnov{}{This JSON serves as an intermediary representation encoding foundational elements, timestamps, relationships, and grammar patterns detected in the video.}

\begin{figure}[h]
  \centering
  \includegraphics[width=\linewidth]{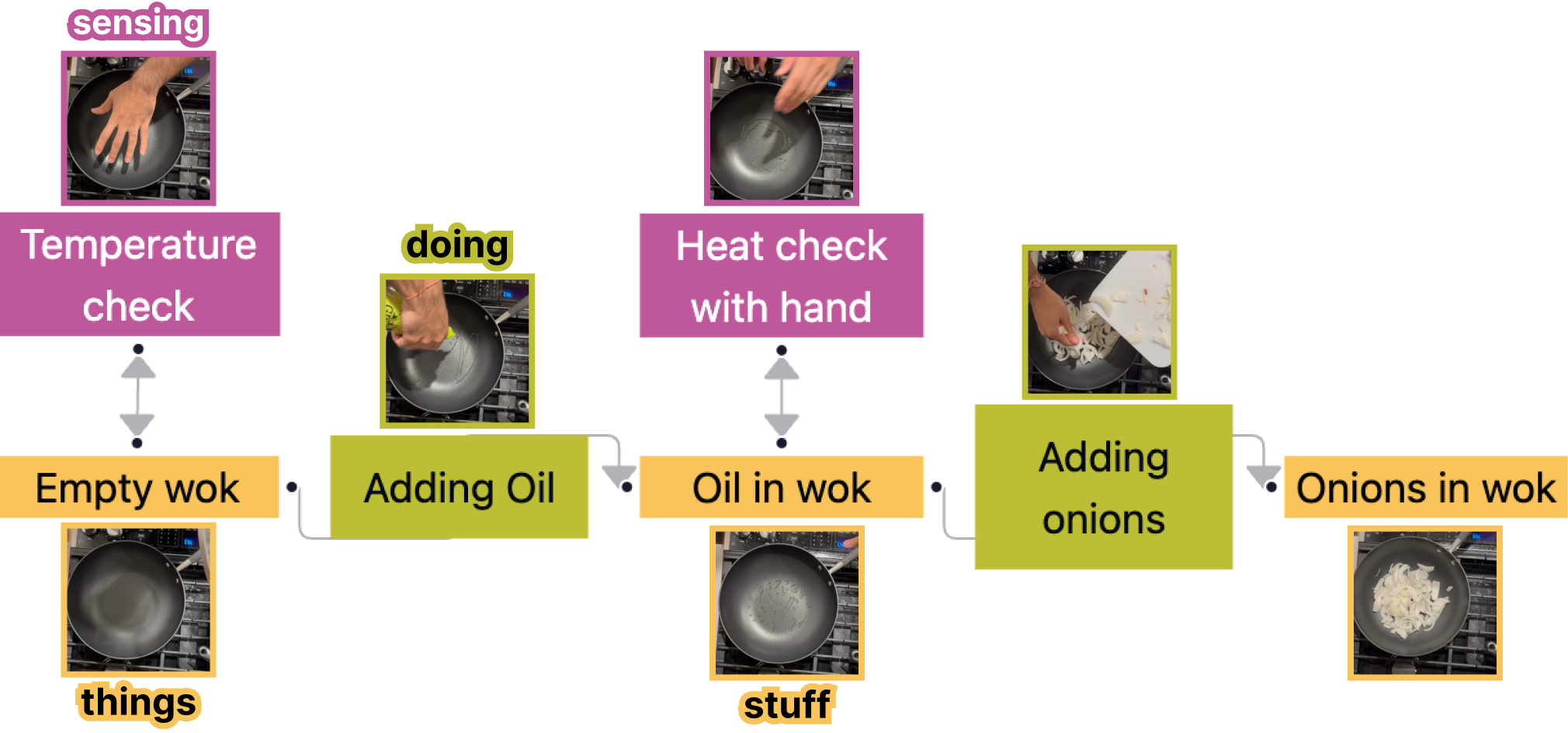}
  \caption{
  The \emph{foundational elements} for our grammar represented in \emph{\interface{}}.
  }
  \label{fig:tool_elements}
  \Description{}
\end{figure}

\subsection{Graph Generation} \label{subsec:genai}

From this generated JSON, we use ReactJS~\cite{react} and the ReactFlow library~\cite{reactflow} to construct a graph \rrnov{}{that visually implements the grammar} (Figure~\ref{fig:workflow}c).
This is a directed graph, where color-coded nodes and edges correspond to \rrnov{}{the grammar's foundational elements}: \emph{stuff} (yellow), \emph{doing} (green), \emph{sensing} (\rrnov{purple}{pink}), and \emph{things} (yellow) as shown in Figure~\ref{fig:tool_elements}, following the definitions by \citet{knight2015making}.
Since \emph{stuff} is explicitly referenced within \emph{things}, the interface combines them into \emph{thing}-nodes.
The graph order is generated as a list of algebraic sequences of \emph{thing} ($T$) and \emph{doing} ($D$) elements, such as $T_1 + D_1 = T_2$.
Reflective loops \rrnov{}{(pink)} are represented as nodes attached to \emph{thing}-nodes, using bi-directional edges to signal \emph{sensing} to inform the following action.
\rrnov{Clicking on nodes or edges allows users to jump to the corresponding video timestamps and navigate the workflow.}{}
Segments \rrnov{}{(purple)} are represented as grouping sequences of \emph{thing}-nodes and a dedicated node for each segment that allows collapsing or expanding the associated segment.
\rrnov{}{Clicking on nodes or edges allows users to jump to the corresponding video timestamps and navigate the workflow.}

\subsection{Refining the Graph} \label{subsec:refine}

The expert can modify the \rrsep{automatically }{}generated graph by editing, deleting, or adding nodes (Figure~\ref{fig:workflow}d), and by appending notes to improve accuracy and clarity (Figure~\ref{fig:workflow}e).
To further enrich the graph with external links, the generated JSON includes references detected in the video, which are supplemented through the Custom Google Search API~\cite{googleCustomSearchAPI}.
Once satisfied with all their changes, they save their modifications back in Google Cloud Storage.
This interactive machine-learning approach~\cite{ramos2020interactive} ensures that the final documentation represents the expert’s workflow and supports varying levels of interpretation and learning.
These editing capabilities enable experts to capture not only the sequence of actions detected in the video but also the rationale, improvisation, and decisions embedded in their workflow.

\subsection{Navigating the Graph} \label{subsec:consuming}

\begin{figure}[h]
  \centering
  \includegraphics[width=\linewidth]{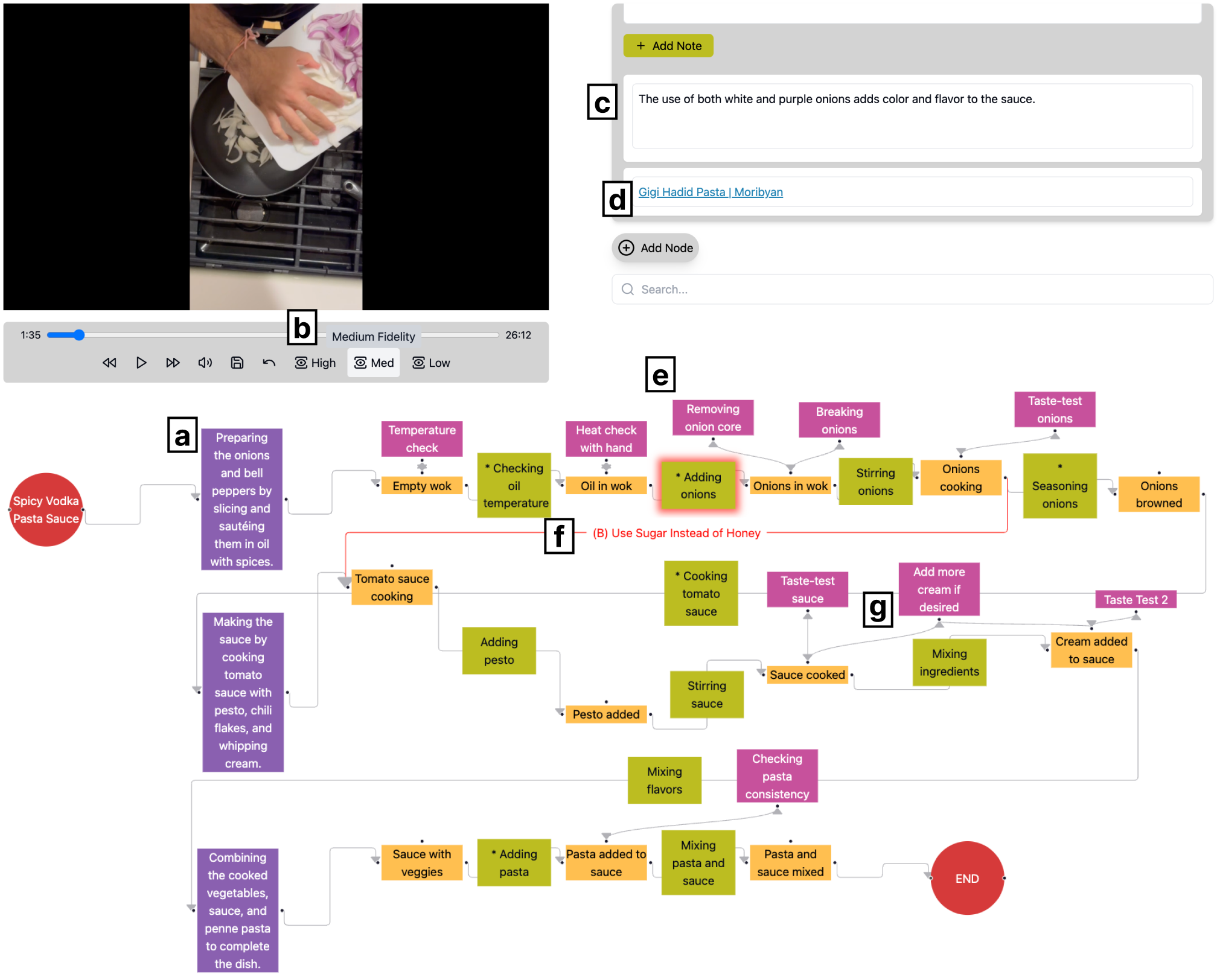}
  \caption{\emph{\interface{}}\rrnov{}{, as shown from the creator perspective,} implements the patterns described within our \grammar{}: \rrnov{(a) \detailblocks{}, (b) \reflective{}, (c) \note{}, (d) \external{}, (e) \branches{}, (f) \revisionloops{}, and (g) \segments{}}{(a) \segments{}, (b) \detailblocks{}, (c) \note{}, (d) \external{}, (e) \reflective{}, (f) \branches{}, and (g) \revisionloops{}}.} 
  \label{fig:interface}
  \Description{}
\end{figure}

Collaborators access the refined graph created by the expert \rrnov{}{(read-only, no edit access)} by visiting the URL and appending `/restore' (Figure~\ref{fig:workflow}f).
The integration of video playback with the expert-modified graph promotes a deeper understanding of expert craft by enabling users to navigate the recording through elements of the grammar, \rrnov{}{visually} represented as nodes and edges \rrnov{}{in the graph}.

\emph{\interface{}} includes features, outlined in Figure~\ref{fig:interface}, that reflect the patterns described in our grammar.
\rrnov{
(a)~The level of detail can be adjusted dynamically using the fidelity controller (\detailblocks{}).
(b)~Reflective moments are visualized as \reflective{}.
(c)~Expert-authored notes (\note{}) provide additional context, (d)~and \external{} provide references to other sources.
(e)~Optional overlays allow users to compare their workflows with those of experts through branching structures (\branches{}).
(f)~\revisionloops{} represent iterative adjustments and corrections, and (g)~the overall workflow is segmented into collapsible groups (\segments{}).
}{
(a)~The overall workflow is segmented into collapsible groups (\segments{}).
(b)~The level of detail can be adjusted dynamically using the fidelity controller (\detailblocks{}).
(c)~Expert-authored notes (\note{}) provide additional context, and 
(d)~\external{} provide references to other sources.
(c)~Reflective moments are visualized as \reflective{}.
(e)~Optional overlays allow users to compare their workflows with those of experts through branching structures (\branches{}).
(f)~\revisionloops{} represent iterative adjustments and corrections.
}
\section{Evaluation} \label{sec:evaluation}

For our grammar, it is critical to consider \rrnov{both the capture of expertise from the documentation creator and sharing of that expertise from the documentation consumer}{reflections from both the creator capturing their expertise and the consumer analyzing the creator's documentation}.
To assess the effectiveness of our \grammar{} in supporting both of these roles \rrnov{}{via \emph{\interface{}}}, we conducted a two-part qualitative evaluation study with seven expert crocheters.
The core research question guiding this study was: \emph{``How effective is our \grammar{} for capturing and sharing craft knowledge \rrnov{between }{within a community of} experts?''}

\subsection{Participants}
\rrnov{Participants (C1-C7) are recruited from the personal network of the authors and each had a minimum of five years of crocheting experience (verified via a background survey conducted at the beginning of the study).}{We recruited seven participants (C1-C7) from our personal networks and prior research engagements in local craft communities. This recruitment method enabled us to sample from a difficult-to-access population of expert crocheters who already have rapport and mutual trust for participating in this study. Each participant had a minimum of five years of crocheting experience (verified via a screening survey at the start of the study) and prior experience with traditional crochet documentation artifacts (e.g., patterns, YouTube tutorials). This ensures participants are between novices and masters, reflecting the middle but wide range of expertise where the \grammar{} can be beneficial.}

\rrnov{\subsection{Evaluation Metric: Grammar Design Guidelines}}{}
\rrnov{
To orient the grammar for craft workflows discussed in the interviews (Section~\ref{sec:interviews}), we observed key patterns across their experiences and identified five design guidelines for designing our evaluation:
\begin{itemize}
    \item \textbf{Generalizability}: Spans diverse craft practices due to their distinct techniques, materials, and documentation methods;
    \item \textbf{Extensibility}: Welcomes additions beyond structures due to the evolving nature of craft;
    \item \textbf{Customizability}: Adapts to the varied expertise and perspectives of each craftsperson and their unique practice;
    \item \textbf{Accessibility}: Supports multiple learning systems due to the non-linear and non-standard paths of craftspeople; and
    \item \textbf{Practicality}: Fits into craft workflows as they already exist without requiring significant changes.
\end{itemize}
}{}


\subsection{Task}
\rrnov{Using a symmetrical study design, we evaluate our grammar from the creator’s and consumer’s perspectives: (1) how effectively experts used the grammar to reflect on and document their own workflows, and (2) how effectively other craftspeople interpreted these representations to understand another expert’s process (interview structure in Supplementary A4). We use a quantitative survey with Likert scales~\cite{likert1932technique} for each part to capture the participant's reactions to the graph's usability after engaging with \emph{\interface{}} as creator and consumer (questions in Supplementary A5).}{We evaluate the grammar from both creator and consumer perspectives: (1) how experts used the grammar to reflect on and document their workflows, and (2) how other craftspeople interpreted these representations to understand the expert’s workflow. All of the participants completed our two-part study conducted via Zoom in April 2025 (evaluation procedure in Supplementary A5).}

\begin{figure}[h!]
  \centering
  \includegraphics[width=\linewidth]{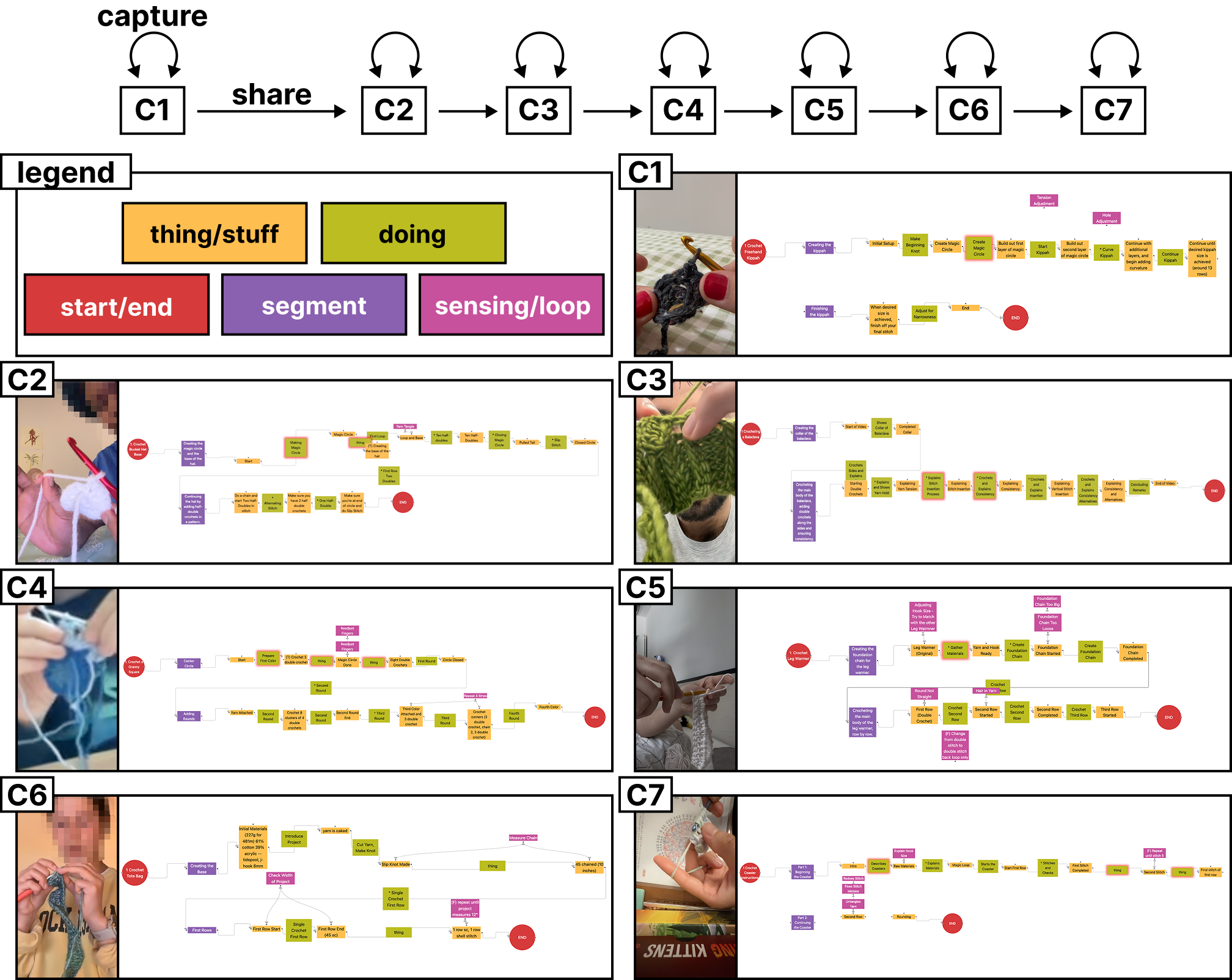}
  \caption{We qualitatively evaluate our \grammar{} with seven expert crocheters, each recording themselves performing a crochet task.}
  \label{fig:evaluation}
  \Description{}
\end{figure}

\rrsep{Part 1: Capture}{}

At least 24 hours before the one-hour Zoom interview, participants recorded, narrated, and shared an unedited 15 to 20 minute video with the authors (Section~\ref{subsec:video}).
To reflect real-world conditions, we incorporated flexibility into the study design by encouraging participants to showcase their expertise within the recorded workflow.
Therefore, we provided instructions to the participants to select a task that highlights a technique or skill they felt best demonstrated their crocheting expertise (the recording did not need to represent a complete project from start to finish).
They then uploaded their video to \emph{\interface{}} before the interview (Sections~\ref{subsec:analyze}, \ref{subsec:genai}).

\rrnov{At the start of the interview}{For the first part of the study}, participants answered demographic and background questions \rrnov{}{via a Qualtrics\footnote{https://www.qualtrics.com/} survey (questions in Supplementary A6)} and \rrnov{then }{}were introduced to \emph{\interface{}}.
\rrnov{Participants could watch their recorded videos alongside automatically-generated workflow graphs along with notes and the other features provided by the grammar.}{Participants could view their recorded videos alongside the automatically-generated graph, which included notes and other features provided by the grammar.}
We asked them to critically evaluate and refine the graph representation by deleting unnecessary elements, adding missing details, editing texts, or reorganizing the graph (Section~\ref{subsec:refine}).
\rrnov{}{Following the task, participants completed a 7-point Likert scale survey~\cite{likert1932technique} via Qualtrics to measure their reactions as documentation creators (survey questions in Supplementary A6).}
Lastly, we conducted \rrnov{qualitative}{a semi-structured} interview\rrnov{s with questions based on the design guidelines}{}\rrsep{ from Section~\ref{sec:grammar}}{} to assess their reflections on the grammar's ability to capture their workflows \rrnov{}{(interview questions in Supplementary A5)}.

\rrsep{Part 2: Sharing}{}

The second part \rrnov{}{of the study} focused on knowledge interpretation\rrnov{}{, aiming to understand how our grammar may support knowledge sharing within a craft community}.
Participants reviewed another participant's refined graph using our interface (Section~\ref{subsec:consuming}).
We used a sequential evaluation approach where Participant $N$ reviewed and discussed Participant $N-1$'s edited graph, as illustrated in Figure~\ref{fig:evaluation}.
For C1, we utilized a \rrnov{workflow }{}graph from a pilot study.
For C2, a software bug resulted in the refined graph not saving so the researchers manually corrected the graph based on the interview recording before the subsequent interview.

Participants spent approximately ten minutes reflecting on the grammar by answering these questions:
(1)~\emph{What do you think the other expert is doing?}
(2)~\emph{\rrsep{Any}{Are there any} particularly interesting decisions they made?}
(3)~\emph{Did anything about the workflow surprise you?}
(4)~\emph{Would you perform any part of this workflow differently?}
\rrnov{}{Following this review, participants completed a second 7-point Likert scale survey to measure their reactions as documentation consumers (survey questions in Supplementary A6).}
At the end of the study, \rrnov{participants responded to qualitative questions oriented towards the}{we conducted another semi-structured interview focused on their} consumption of \rrnov{another expert's}{the other participant's} documentation artifact \rrnov{}{(interview questions in Supplementary A5)}.
\rrnov{All participants had prior experience with traditional crochet documentation artifacts (e.g., patterns, YouTube tutorials).}{}

\subsection{Data Analysis}

Interviews were transcribed using \rrsep{}{a local instance of} Whisper's pre-trained large-v2 model~\cite{radford2022whisper}.
We follow a six-step hybrid inductive-deductive coding and thematic analysis process for analyzing these transcripts \rrnov{}{(including the interview responses)} as outlined in \citet{fereday2006demonstrating} and similarly applied in recent CSCW literature~\cite{kim2024occupational, liaqat2019social, he2022help}.
(1)~We developed an initial codebook \rrnov{based on design guidelines}{based on the patterns observed in our expert interviews (Section~\ref{sec:interviews})}\rrsep{ detailed in Section~\ref{sec:grammar}}{}.
(2)~We evaluated the reliability of these initial codes by applying these codes to two of the interviews.
This surfaced inductive codes that did not fit within the initial framework, prompting revisions to the codebook (44 total codes).
(3)~We identified themes by creating memos to process the raw data and interview responses from each transcript.
(4)~Then, we systematically coded all interviews, guided (but not constrained) by the preliminary codebook as additional inductive codes emerged \rrsep{}{until thematic saturation was reached} (77 total codes).
(5)~We iteratively developed themes by discussing patterns across the codes and arrived at the final set of 4 themes that demonstrated saturation across all interviews.
(6)~Finally, we refined these themes through discussions during which disagreements were resolved and alternative interpretations were carefully considered by the team of researchers from various disciplines (e.g., human-computer interaction, science and technology studies).

\subsection{Findings}

\rrsep{We structure the results from this study around four themes that reflect the grammar’s utility and tie back to our design guidelines from Section~\ref{sec:grammar}.}{We structure the results of this study around four themes that reflect the grammar’s utility.}

\subsubsection{Level of Detail: Sharing Knowledge with Diverse Consumers} \label{eval:detail}

All participants emphasized the importance of supporting varying levels of detail\rrnov{ in the grammar}{}, highlighting this as a necessary feature \rrnov{}{of the grammar} for accommodating different audiences and project complexities\rrnov{(related design guidelines: \textbf{customizability}, \textbf{accessibility})}{}.
Participants acknowledge the grammar's ability to support dynamically detailed documentation, with C3 asking, \emph{``Are you trying to explain this to someone who's never done a magic circle before? Then you really need to go deep in detail.''}\rrnov{ (\textbf{extensibility}).}{}
C1 highlighted how the grammar allowed them to selectively spotlight the most critical steps and preview the craft workflow, noting, \emph{``I have to decide here the level of detail that I want, because I think more detail might confuse people''} and \emph{``it's just hard to visualize how that craft is made without getting like the one, two, three high-level main steps at first... this is something that I 100\% resonate with because anytime I try to understand anything, I want to know the synopsis first.''}
This suggests this variation is helpful not only for consumers of varying expertise but also for a single (even expert) consumer over the course of developing their knowledge and practice.
C2 shared how the grammar prompted personal reflection on the documentation artifact and how greater specificity could improve it to better support others' understanding of the workflow: \emph{``I didn't mention how many rows... that's definitely something that I should mention next time.''}
Across participants, the grammar’s ability to adjust documentation granularity was seen as essential for capturing and sharing craft workflows effectively.

\subsubsection{Flexibility: Supporting Tacit Knowledge and Improvisation} \label{eval:flexibility}

All participants emphasized the importance of flexibility in the grammar's structure to accommodate diverse and improvisational craft workflows\rrnov{ (\textbf{customizability}, \textbf{extensibility})}{}.
Four participants described their practice as guided more by freehanding than by formal patterns, highlighting how the grammar supported their partial and evolving workflows in contrast to rigid sequences of traditional documentation methods. 
As C1 reflected, \emph{``I'm not a great person for [creating documentation] because I don't know the proper language for these things,''} adding, \emph{``I would just try to do this without reading a pattern and it wouldn't come out right, which is what I know happens. But yeah, that's the only way that I can craft.''}\rrnov{ (\textbf{extensibility}).}{}
Conversely, C2 and C5 noted some challenges with freehanding specific patterns, as C5 observed, \emph{``a granny square usually follows a specific pattern,''} and C2 echoed that it would be \emph{``very difficult to freehand a complete granny square.''}
Beyond replicating formal patterns, participants like C6 valued the grammar’s ability to support unstructured workflows and similarly non-linear representations for documenting their practice, stating, \emph{``I actually don't mind it being kind of disorganized because I think that's the flow I've had forever and what I'm really used to.''}\rrnov{ (\textbf{customizability}).}{}
C6 also shared how the grammar can capture the \emph{``revision history of `This is what I originally had, this is how I fixed it.' And then maybe, `When I tried to fix it, it actually didn't work,'''} highlighting how the grammar matches the non-linear nature of expert workflows.
C3 emphasized the importance of representing and consolidating variations in craft workflows, suggesting how the grammar \emph{``could be a great way to aggregate or summarize different ways of doing things because there's a million different ways to make a bucket hat or a beanie. And you can do it in the round, you can do it flat, and you can sew it.... There's so many decisions you can make and... [the grammar] could be an interesting way to aggregate [those decisions].'''}\rrnov{ (\textbf{extensibility}).}{}
Together, these reflections illustrate how flexibility within the grammar was seen as essential for capturing both the diversity and the adaptive nature of expert craft practice.

\subsubsection{Capturing vs. Sharing: Bridging Craft Knowledge and Expertise Sharing} \label{eval:grammar}

Five participants reflected on how the grammar supported different goals: capturing workflows for personal use versus structuring them for teaching others\rrnov{ (\textbf{customizability}, \textbf{practicality})}{}. 
All participants expressed enthusiasm for sharing their craft knowledge, with C3 stating, \emph{``I love to teach other people how to crochet.''} 
While the grammar enabled recording intuitive, embodied processes, participants noted that producing representations intended for others often required additional restructuring.
C1 described adjusting their edits depending on the audience, explaining, \emph{``If I'm making [documentation] for somebody else, I would have edited the graph the way that I did.... The way that I work is just not the most efficient and so if I'm explaining it to somebody, I would fix some things.''}\rrnov{ (\textbf{practicality}).}{}
Similarly, C5 highlighted the grammar's potential for supporting tutorial creation, noting, \emph{``If I ever want to make like a tutorial video one day, then it would be pretty helpful.''}\rrnov{ (\textbf{customizability}).}{}
C7 compared our grammar to existing documentation platforms like \emph{Ravelry}, describing how they often \emph{``Frankenstein''} patterns together, highlighting how experts also adapt and remix others’ documentation to document their own workflows\rrnov{ (\textbf{customizability}, \textbf{generalizability})}{}.
C4 also shared how the grammar supported their learning of nuances and techniques from C3's documented workflow: \emph{``I think that like the overall explanation of like, here's how you find the head of a stitch and here's where you insert your hook... in between that head of a stitch and the next.... That was a useful explanation that I will probably remember if I need to do that in the future.''}\rrnov{(\textbf{accessibility}).}{}
C4 acknowledged that while they did not normally document their workflows for others, the grammar streamlines creating documentation artifacts for them to reference aspects of their older projects: \emph{``[the grammar is] super helpful for... reconstruct[ing]''} previous projects\rrnov{ (\textbf{customizability})}{}.
These reflections illustrate how the participants valued adaptability in the grammar to document personalized, unstructured workflows, while still providing enough structure to support their knowledge development and sharing across craftspeople.

\begin{figure}[h!]
  \centering
  \includegraphics[width=\linewidth]{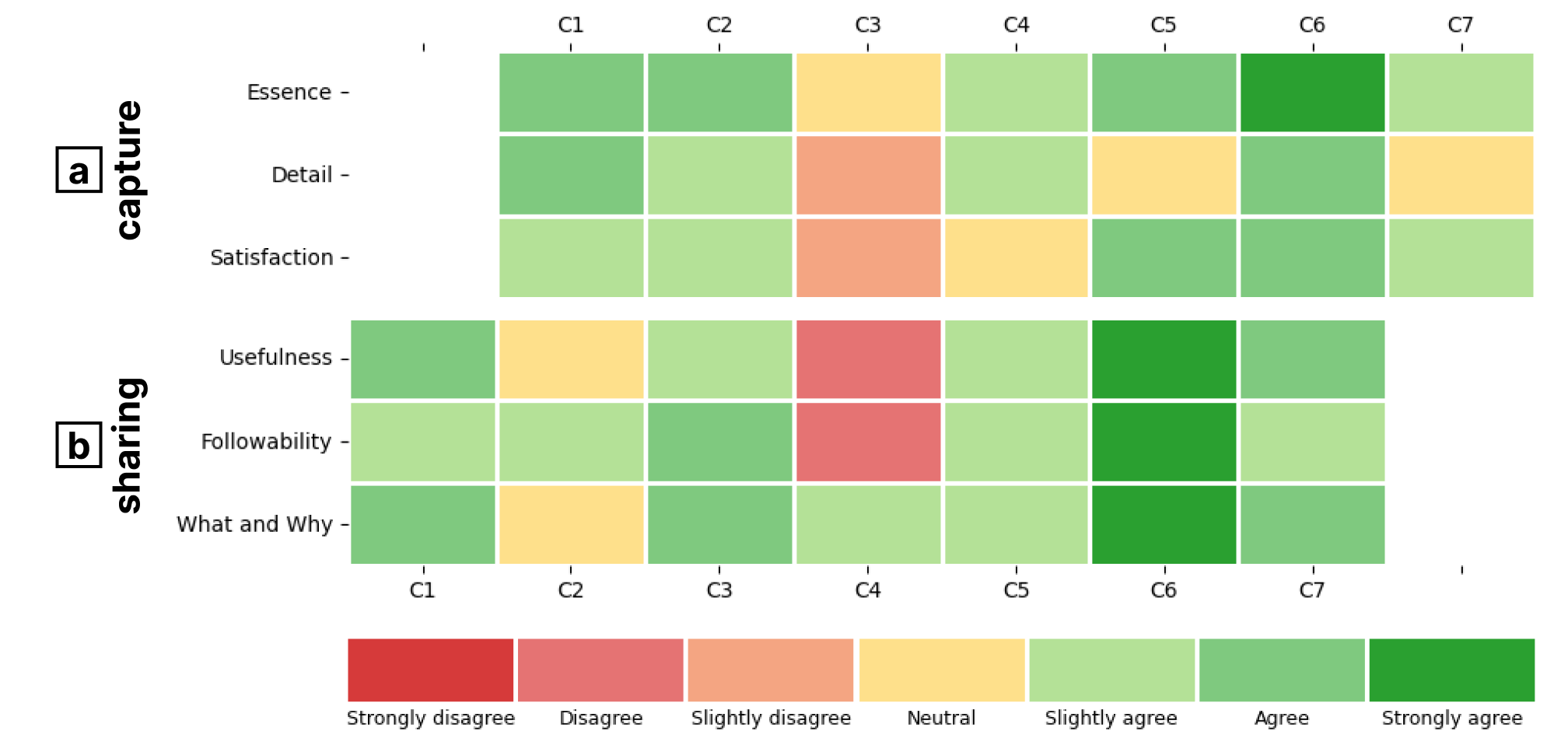}
  \caption{Participant responses to survey questions evaluating the grammar’s effectiveness \rrnov{}{(as implemented in \emph{\interface{}})} in (a)~capturing their own workflow and (b)~sharing knowledge by analyzing another participant's graph.}
  \label{fig:quant}
  \Description{}
\end{figure}

As shown in Figure~\ref{fig:quant}, the participants’ perceptions \rrnov{of the documentation quality using the grammar}{of the grammar's effectiveness as reflected through the graphs} were aligned between the creators and consumers in the study.
For example, C6 agreed that the graph captures their own workflow, and C7 (who reviewed C6's graph) was able to understand C6's workflow.
Conversely, C3 showed dissatisfaction in the grammar's ability to capture their workflow, and C4 (who reviewed C3's graph) \rrsep{indicated that the workflow}{also indicated that the graph} was unclear.

\subsubsection{\rrnov{What is Lost?: Addressing Challenges of Documentation}{What is Lost: Limits of Documentation}} \label{eval:whatislost}

Four participants discussed challenges of using the grammar to translate their workflow into a structured representation \rrnov{(\textbf{accessibility}, \textbf{practicality})}{similar to the graph}.
As C1 admitted, \emph{``I'm not confident that I still understand the difference between the yellow [things] and the green [actions],''} a sentiment echoed by others as they tried to map familiar processes into the grammar’s structure.
C2 reflected on the difficulty of aligning their natural narration with the grammar’s demands, stating, \emph{``Sometimes I don't say what's happening,''} highlighting a tension between spontaneous craft practice and externalizing knowledge\rrnov{ (\textbf{practicality})}{}.
Participants such as C6 also described the graph as \emph{``a little bit awkward just because I'm so used to reading [patterns] like a book or on a page,''} detailing the challenges in transitioning from familiar documentation conventions to a new documentation framework\rrnov{ (\textbf{accessibility})}{}.
These reflections suggest that while the grammar offers new affordances for capturing and sharing workflows, it also introduces unfamiliar structures that present a key design tension for documenting craft practices.

While the experts acknowledged the utility of the grammar, all participants also reflected on what remains difficult to capture through structured documentation, particularly the embodied and relational aspects of craft practice\rrnov{ (\textbf{accessibility}, \textbf{practicality})}{}.
C4 emphasized the limits of formal representation when describing physical technique, noting, \emph{``I just like feel [it] in my body at this point... I have trouble putting it into words.''}
They described elements like yarn tension and stitch anatomy as \emph{``not really something that you see in patterns or written instructions''} and found such knowledge \emph{``a little unintuitive in this graph.''}\rrnov{ (\textbf{accessibility})}.
While the graph served as the visual interface \rrnov{}{for the grammar}, these reflections gesture toward deeper challenges in representing sensory and embodied knowledge within \rrnov{}{any} externalized formats.
Others echoed concerns about the absence of real-time responsiveness, with C1 calling this \emph{``an unavoidable trade-off''} and C4 expressing that they would not \emph{``trust it to get everything perfect and the small details of... the anatomy of a stitch,''} further highlighting the underlying limits of documentation for capturing tacit knowledge\rrnov{ (\textbf{practicality})}{}.
Additionally, C2 admitted, \emph{``I think I just get confused by what the nodes do''}, which suggests that the \rrnov{structured representations require}{graph's representation of the grammar requires} clear on-boarding to increase the grammar's interpretability\rrnov{ (\textbf{accessibility})}{}.
These reflections illustrate that while the grammar supports structured capture and sharing, it cannot substitute the tacit, sensory, and interactive dimensions of expert craft.
\section{Discussion}

How does an expert craftsperson capture and share their \rrnov{knowledge}{tacit knowledge, improvisational actions, and situated adaptations} within an evolving community of practitioners, materials, and tools?
Conversely, how can that \rrnov{same community collaborate to support the evolution of a collective archive of craft knowledge}{community of craftspeople collaborate to support an evolving archive of craft knowledge}?
\rrnov{We address these questions through the findings and observations of this work. In doing so we contribute additional support for existing theories, identify opportunities for further exploration, and discuss limits of computer-supported collaboration through the lens of our findings.}{From our findings, we observe four core tensions that shape how computational systems can support craft documentation and knowledge sharing: personal and shareable documentation; fragmented and discoverable expertise; linear and iterative practices; and data privacy and ownership.}

\subsection{\rrnov{Craft Knowledge and Expertise Sharing}{Personal and Shareable Documentation}}

\rrnov{We}{From our evaluation study, we} observed a distinction between personal documentation and shareable documentation intended for broader consumption (Section~\ref{eval:flexibility}).
Personal documentation\rrnov{}{, described as \emph{``inefficient''} by C1,} often manifests as unstructured records that make sense to their creators but remain incoherent to others~\cite{meiklejohn2024design}. 
These \emph{scrappy} representations (e.g., quick sketches, shorthand notes, personalized diagrams) \rrnov{are often}{may be} meaningful to the creator but \rrnov{}{are} difficult for others to interpret or extend, presenting a tension when developing archives of knowledge intended for broader use.
Our grammar served as a method for translating between these forms: \rrnov{a way to scaffold shareable documentation while supporting unrestricted workflows that do not require the expert to constantly consider the interpretability of their documentation}{capturing the improvisational adjustments critical for personal documentation and then organizing them into navigable graphs that enabled other experts to learn specific techniques (Section~\ref{eval:grammar})}.

This translation between personal and shareable documentation becomes particularly important in the context of how experts collaborate to learn from and contribute to communal knowledge archives.
\rrnov{Craftspeople leverage}{Participants in our study leveraged} a range of learning pathways, from YouTube tutorials and local workshops to informal exchanges with peers, highlighting how craftspeople seek knowledge and learn to navigate the uncertainties of their practice (Sections~\ref{interviews:limitations},~\ref{eval:flexibility}).
However, the challenge is developing documentation systems that capture both craft expertise and accommodate revision, layering, and collaborative reinterpretation over time.
This extends beyond craft; community-driven knowledge platforms such as Wikipedia~\cite{wikipedia} have long grappled with how to structure bodies of knowledge that are both evolving and collaboratively authored~\cite{hill2020wikipedia}.
For these platforms, the goal is not for a single documentation creator to produce perfect, immutable knowledge \rrnov{}{for others to \emph{``reconstruct''} (C4)} but to create the scaffolding for knowledge to be iteratively extended, revised, and linked.
Without such a structure, knowledge archives risk becoming \emph{stale}, or frozen in a single creator’s context and time (as often seen in YouTube videos), disconnected from the evolving needs and contexts of other creators.
Our grammar's design enables interoperability between disparate, idiosyncratic forms of expertise, transforming them into shareable and extensible documentation artifacts (Section~\ref{eval:grammar}).
The grammar is designed to enable distributed authorship and iterative knowledge building, using patterns to emphasize the collaborative and evolving nature of expert knowledge.

With documentation that evolves over time and across contributors, the grammar supports the visibility of expert deviations that emerge as experts bend and break the \emph{rules} of their practice~\cite{abbott2014system, berliner2009thinking}.
However, creating a knowledge archive of individual improvisations requires more than cataloging variations; it requires contextual framing that makes these improvisations legible and applicable to others within the community (Section~\ref{eval:flexibility}).
Our grammar helps surface and structure improvisations in expert workflows by offering a shared language for describing deviations and their motivations. 
To support the sharing of this knowledge, we argue for external representations~\cite{suthers2003deictic, fischer2005knowledge} that augment unstructured video by guiding others in navigating and interpreting improvisations within their original contexts.
A knowledge archive of individual improvisations, contextualized for others to transfer into their own practice, shifts collective knowledge toward a shared but situational understanding of when, why, and how \emph{rules} are bent~\cite{abbott2014system}.

\subsection{\rrnov{Frankenstein-ed Knowledge Archives}{Fragmented and Discoverable Expertise}}

\rrnov{In this section, we discuss the creation of \emph{Frankenstein}-ed knowledge archives, where expertise draws on a branches of fragmented sources of knowledge, rather than a single source. As experts compile patterns, examples, and partial instructions into a single documentation artifact, new challenges emerge around the search for expert knowledge and variations in expertise.}{As experts compile patterns, examples, and partial instructions into what C7 described as \emph{Frankenstein}-ed knowledge archives, new challenges emerge around discovering and accessing expert knowledge.}
\rrnov{Frankenstein-ed knowledge archives often obscure the retrieval and recognition of expert knowledge, especially the insights embedded in practice that are rarely made explicit.
Such knowledge is often implied through adaptation, improvisation, or deviation, yet remains difficult to locate and access within documentation systems (Section~\ref{eval:whatislost}). 
While documentation may be abundant, its utility is constrained by the lack of indexing mechanisms that accommodate ambiguity, interpretation, and context as discussed by \citet{torrey2009learning}.}{A critical component of navigating these fragmented archives is \emph{search}. Craftspeople engage in iterative cycles of searching for knowledge across YouTube tutorials, forum discussions, and pattern libraries. However, certain types of expert knowledge remain difficult to articulate in text, making them difficult to locate and access due to the lack of indexing mechanisms that accommodate ambiguity, interpretation, and context as discussed by \citet{torrey2009learning} (Section~\ref{eval:whatislost}).}
In other domains, parallel efforts such as SplatOverflow~\cite{kwatra2024splatoverflow} offer a path forward by developing methods that bridge digital representations and physical objects to make hardware issues searchable in ways similar to software troubleshooting on StackOverflow~\cite{stackoverflow}.
These cases underscore the need for documentation systems that enable flexible, context-aware querying of knowledge that inherently resist formalization.

In this work, we highlight the importance of treating documentation not as a static source of ground truth but as a flexible, collaborative artifact \rrsep{of knowledge}{--- a core part of the evolving archive and knowledge infrastructure of craft communities}.
Rather than documenting workflows with an authoritative structure, our grammar supports the reinterpretation and remixing of documentation by providing a common method that links otherwise disparate representations (e.g., sketches, notes, photos) into a coherent and navigable workflow.
This approach to documentation serves less as rigid instruction and more as a provocation for improvisation and reinterpretation (Section~\ref{interviews:improv}).
This perspective reframes documentation as part of an ongoing exchange between \rrsep{an }{}expert and \rrsep{the }{}community where practitioners actively reshape and contribute back to collective knowledge \rrsep{}{and practice}.
As \citet{ackerman2013sharing} discuss, documentation artifacts should stimulate the socialization and internalization of knowledge, further emphasizing the collaborative dimension of expert documentation.
Rather than viewing Frankenstein-ed knowledge archives as a sign of disorder, they represent a vital mode of participation in the exchange of knowledge between experts and their community \rrsep{to advance the collective understanding of the practice}{that both advances and evolves craft practice, supporting its continuity and innovation over time}.

\rrnov{
\subsection{Challenges in Creating Documentation} 
}{}

\rrnov{
There is an inherent tension between computational tools and craft practice, which highlights the limits of computational systems in capturing the full texture of real-world skill and interactions.
As we consider how expertise is shared within communities, such limitations become especially clear: computing typically struggles to enable the live feedback essential to pedagogy and capture the full extent of what constitutes as expertise~\cite{corbett1997intelligent, kim2017mosaic}.
}{}

\rrnov{
Apprenticeship remains the gold standard for knowledge sharing, grounded in intimate, embodied collaboration between expert and learner~\cite{yarmand2024d, lave1991situated}.
As discussed by research on the timing of feedback~\cite{jane2024feedback, cheng2020critique}, these live interactions during the expert action enables a dynamic responsiveness that post-hoc documentation artifacts cannot replace.
For example, working side-by-side with students enables them to observe the nuanced decisions, ask questions, and refine their understanding in real-time~\cite{thoravi2019loki} (Section~\ref{interviews:limitations}).
While computing may never fully reproduce this co-presence, our grammar scaffolds aspects of this pedagogical exchange and collaboration by showcasing expert thought processes and the variability of their practices. 
In this way, documentation augments collective knowledge, capturing and sharing expertise to support communities of practice.
}{}

\rrnov{
Although grammars provide flexible pathways to describe and share\rrsep{ the}{} knowledge within expert workflows, any computational system representing knowledge remains partial; as \citet{korzybski1931non} reminds us, \emph{``the map is not the territory.''}
Our grammar, like other documentation approaches, still leans on visual and structural representations which struggle to convey the full sensory and tactile aspects of many skilled practices (e.g., glassblowing, ceramics)~\cite{mueller2003marvel}.
However, by creating methods to share and interpret expert decision-making and improvisations, our approach shifts documentation closer to lived practice, not by attempting to replace embodied learning but by inviting engagement with the uncertainties and intuitions that underscore expertise.
While the map may not be the territory, our grammar charts pathways toward shared understanding and deeper collective knowledge, reaffirming that \emph{(de)composing craft} into non-linear, dynamic documentation artifacts further supports a community that is collaborating, adapting, and growing together.
}{}

\subsection{\rrnov{Grammars to Map Expertise}{Linear and Iterative Practices}}

Across our studies, craftspeople shared nuances of their workflows that characterize the essence of their expertise but are rarely captured in conventional documentation.
These nuances driven by tacit knowledge often exist at the margins of workflows, making them difficult to observe, describe, and share~\cite{polanyi1967tacit}.
However, we observed that it is in these moments where expertise manifests as craftspeople respond to material properties, reconfigure their tools, and revise their plans to minimize the \emph{risk} that Pye describes as inherent to craftsmanship~\cite{pye1995nature} (Section~\ref{interviews:improv}).

Knight and Stiny's theory of \emph{Making Grammars}~\cite{knight2015making} offers a framework for capturing how sequential operations transform initial materials into final forms.
Our findings challenge the unidirectional, linear approach dominant in documentation practices.
Rather than describing workflows as a sequence of instructions, we found that craftspeople emphasize the prototyping, mistakes, and iterations that led to the final form (as well as the pedagogical benefits beyond the artifact discussed in Section~\ref{interviews:learning}).
\rrnov{When materials crack, tools falter, or plans go awry, experts respond with improvisation, treating \emph{repair}~\cite{jackson2014rethinking} in workflows not as a detour but as an integral part of the creative trajectory~\cite{goveia2022portfolio} (Section~\ref{interviews:improv}, \ref{eval:flexibility}).}{The participants in our study valued documenting this revision history to capture not just successful outcomes, but also the failed attempts and lessons that shape their practice (Section~\ref{eval:flexibility}). When materials crack, tools falter, or plans go awry, experts respond with improvisation, treating repair~\cite{jackson2014rethinking} in workflows not as a detour but as an integral part of the creative trajectory~\cite{goveia2022portfolio}.}
In this context, expertise emerges not only in generating form, but also in effectively responding to unpredictable contexts.

As discussed \rrnov{}{by participants} in our evaluation study (Section~\ref{eval:whatislost}\rrsep{}{)}, externalized representations have limitations for capturing expert knowledge~\cite{suthers2003deictic}.
\rrnov{The}{C1 characterized this as an \emph{``unavoidable trade-off,''} acknowledging that while documentation can guide understanding, it cannot fully capture the} tactility of yarn tension or the intuitive shaping of a stitch\rrnov{ remain difficult to articulate and communicate}{}.
\rrnov{While notation systems like Labanotation~\cite{guest1977labanotation} and Gilbreths' motion studies~\cite{gilbreth1917applied} attempt to encode expertise and intent in other domains, they too struggle to capture the rich nuance of expertise.}{}
Our grammar opens pathways to express expert thinking as it emerges through the interplay of body, materials, and tools \rrnov{}{by documenting the reflective loops, branches, and revisions that reveal how experts navigate uncertainty}.
The resulting documentation artifacts highlight how an expert's plan may not match the final execution as the workflow is shaped by the friction of the world rather than its idealized smoothness.

\rrnov{This approach extends the notion of \emph{breakdown} as a generative site of knowledge~\cite{jackson2014breakdown} and how documentation can capture both the procedure and how expert decisions unfolded in response to friction, failure, and decay~\cite{ingold2009Textility}.}{}
\rrnov{In this way, documentation becomes bi-directional and cyclical, charting not just the forward path of making but the inverse and iterative processes of breakdown, reflection, and repair through which final forms emerge.
This reframing of documenting expertise aligns with how tacit knowledge develops in real-world scenarios~\cite{ingold2013making} (e.g., when tools fail, materials resist, or plans unravel) and is demonstrated through \emph{reflection-in-action}~\cite{schon1979reflective}.
While conventional documentation often focuses on ideal plans and perfect forms, our grammar captures expert practice as it actually unfolds: full of dead ends, reversals, and tangents.}{In this way, documentation becomes bi-directional, cyclical, and relational, charting not just the forward path of making but the inverse, iterative, and interconnected processes of breakdown, reflection, and repair through which final forms emerge~\cite{ingold2013making, schon1979reflective, bennett2020vibrant}. This re-framing of documenting expert practice with its dead ends, reversals, and tangents offers pathways toward the development of new computational infrastructures that better support the distinct modes of seeing, thinking, and doing that are essential to craft practice.}


\subsection{\rrnov{}{Data Privacy and Ownership}}

\rrsep{}{
Integrating third-party services (here, Google Gemini) into \emph{\interface{}} to facilitate knowledge sharing between craftspeople raises concerns about data privacy and ethics. We received informed consent to share participant videos with Google Gemini prior to the study. Due to concerns about videos being used as training data for AI~\cite{baack2025towards}, we allowed and encouraged participants to adjust their approach to recording their narrated videos. For example, participants such as C1, C5, and C7 recorded the crocheting video from camera angles that excluded their face. To circumvent data being shared with third-party services at all, future interfaces similar to \emph{\interface{}} could use offline MLLMs such as the models integrated with Ollama~\cite{ollama2025} that process videos without sending data to third-party services.
}

\rrsep{}{
Another ethical consideration is the sharing of expert craftspeople's intellectual property. We envision the grammar being primarily used to create documentation for community knowledge sharing, a practice that was already familiar and common among study participants. All participants disclosed prior experiences of sharing knowledge. C2, more specifically, often posts their crocheting videos on Instagram. By demonstrating the \grammar{} with an interface that could be modified before sharing, participants decide what aspects of their video are shared/withheld. This pattern of selective sharing and withholding is consistent with existing norms of exchange in craft communities. We hope future work in developing documentation interfaces preserves this autonomy for craftspeople to selectively determine what knowledge is captured and distributed.
}

\section{Conclusion}

\rrnov{In this paper, we demonstrate how an \grammar{} can support the capture and sharing of expert knowledge in non-linear craft workflows, highlighting the moments of reflection, improvisation, and adaptation that underscore expertise.
To demonstrate how the grammar works in practice, we developed \emph{\interface{}}, a documentation interface that uses the grammar and an MLLM to process unstructured crafting videos and generate structured graphs emphasizing improvisational actions from the workflow.
Through a two-part study with expert crocheters using \emph{\interface{}}, participants were able to edit and reflect on their workflow's graph and then critically engage with another expert's graph.
The experts reflected on how the grammar surfaces tacit decisions and makes visible the otherwise invisible work behind expert practice.
By introducing a more expressive and flexible documentation methodology, this work builds new pathways for capturing and sharing expert knowledge across time, space, and skill levels, while fostering collaboration and distributed contribution within communities of practice.}{}

\rrnov{}{
In this paper, we have explored how expert craftspeople capture and share tacit knowledge, improvisational actions, and situated adaptations within evolving communities of practice.
Through interviews with expert craftspeople across diverse domains, we developed an \grammar{} that documents the non-linear, iterative, and improvisational nature of craft workflows.
From our evaluation of the grammar with expert crocheters, we learned how experts navigate documentation to preserve a dialogue between craftsperson, material, and community. This dialogue emerges in distinct yet interconnected ways: through \reflective{} where experts respond to material feedback, \branches{} where they deviate from established procedures, \external{} where they reference other knowledge sources, and \revisionloops{} where they iterate and repair.
In addition, practitioners extend and articulate their practice with \detailblocks{}, \segments{}, and \note{}.
With this framing, we argue for an expanded understanding of documentation tools for craft practices, not as static repositories of perfect instructions, but as dynamic scaffolding for craft knowledge to be extended, revised, and linked across distributed communities of craftspeople.
}

\rrnov{}{
As with any computational system representing knowledge, grammars remain partial representations; \citet{korzybski1931non} reminds us, \emph{``the map is not the territory.''}
Documentation serves to augment (not substitute) the hands-on practice and direct material engagement essential to the learning and evolution of practice in craft communities. 
Nevertheless, computational tools such as \emph{\interface{}} can make modest but meaningful contributions to these wider practices.
Our evaluation revealed opportunities for future work: exploring interfaces more familiar to specific craft communities, investigating how the grammar might support practitioners with varied expertise levels, and studying how documentation artifacts evolve through revising and remixing over time.
While our grammar may not capture the full territory of craft practices, it demonstrates how computational tools and techniques, suitably and sensitively deployed, can better support the distinctive conditions of sharing and learning found in craft communities.
By formalizing patterns of improvisation, iteration, and repair, we can help make tacit knowledge visible and shareable, supporting craft communities in building collaborative and evolving archives that preserve not just final outcomes but the expertise embedded in their creation.
}

\bibliographystyle{ACM-Reference-Format}
\bibliography{thijs-papers, references}


\end{document}